\documentclass[12pt,preprint]{aastex}
\usepackage{graphicx}

\begin{document}

\title{Photoevaporation of Clumps in Photodissociation Regions}

\author{U.Gorti and D.Hollenbach}

\affil{NASA Ames Research Center, Moffett Field, CA}
%
\begin{abstract}
We present the results of an investigation of the effects of Far Ultraviolet
(FUV) radiation ($6.0 eV < h\nu < 13.6 eV$) from hot early type OB stars on
clumps in star-forming molecular clouds. Clumps in FUV-illuminated regions
(or photodissociation regions or PDRs) undergo external heating and
photodissociation as they are exposed to the FUV field, resulting in a 
loss of cold, molecular clump mass as it is converted to warm atomic gas.
 The heating, if rapid, creates strong photoevaporative mass flows off
the clump surfaces, and drives shocks into the clumps, compressing them to
high densities. The clumps  lose mass on relatively short timescales. The
evolution of an individual clump is found to be sensitive to three dimensionless
parameters: $\eta_{c0}$, the ratio of the initial column density of the clump
to the column $N_0 \sim 10^{21}$ cm$^{-2}$ of a warm FUV-heated surface region;
$\nu$, the ratio of the sound speed in the heated surface to that in the cold
clump material; and $t_{FUV}/t_{c}$, the ratio of the ``turn-on time'' $t_{FUV}$ 
of the heating flux on a clump to its initial sound crossing-time $t_{c}$.
The evolution also depends on whether a confining interclump medium exists, or
whether the interclump region has negligible pressure, as is the case for
turbulence-generated clumps.  In this paper, we use
spherical 1-D numerical hydrodynamic models as well as approximate analytical
models to study the dependence of clump photoevaporation on the physical
parameters of the clump, and to derive the dynamical evolution, mass loss rates
and photoevaporative timescales of a clump for a variety of astrophysical situations.
Turbulent clumps evolve so that their column densities are equal to a critical
value determined by the local FUV field, and typically have short photoevaporation
timescales, $\sim 10^{4-5}$ years for a 1 M$_{\odot}$ clump in a typical star-forming
region ($\eta_{c0}=10, \nu=10$). 
 Clumps with insufficient magnetic pressure support, and in strong
FUV fields may be driven to collapse by the compressional effect of
converging shock waves. 
We also estimate the rocket effect on photoevaporating clumps and find that it is
significant only for the smallest clumps, with sizes much less than the extent of
the PDR itself. Clumps that are confined by an interclump medium may either get completely
photoevaporated, or may preserve a shielded core with a warm, dissociated, protective
shell that absorbs the incident FUV flux. 
We compare our results with observations of some well-studied PDRs: the
Orion Bar, M17SW, NGC 2023 and the Rosette Nebula. The data are
consistent with both interpretations of clump origin, turbulence and pressure
confinement, with a slight indication for favouring the turbulent model for clumps
over pressure-confined clumps.
\end{abstract}

%
\section{Introduction} Young massive OB stars significantly influence the
structure, dynamics, chemistry and thermal balance of their associated
molecular clouds through the impact of their ultraviolet photons.
Their extreme ultraviolet photons (EUV, $h\nu >$ 13.6 eV) ionize the gas
immediately surrounding them, forming \ion{H}{2} regions.  Far ultraviolet
photons (FUV, 6 eV $< h\nu<$ 13.6 eV) dissociate molecular gas beyond the
\ion{H}{2} region, creating a photodissociation region or  PDR
(Hollenbach \& Tielens 1999).
PDRs are ubiquitous: FUV photons dominate the gas heating in PDRs and affect
the chemistry and physics of gas over a large fraction of the volume and
mass of molecular clouds. The study of interactions between FUV radiation
and molecular cloud gas is therefore important in understanding molecular
cloud evolution and the feedback between massive star formation and
subsequent star formation in molecular clouds.

Observations of PDRs compared with theoretical models probe the physical
conditions in star-forming regions through dust and gas emission mainly in
the infrared (IR) wavelength region of the spectrum. PDRs are the source of
most of the IR emission in the Galaxy. Dust grains and large carbonaceous
molecules such as polycyclic aromatic hydrocarbons (PAHs) absorb radiation
from the stars, and re-radiate this energy flux in the infrared, producing
a continuum with solid state and PAH spectral signatures. FUV photons incident
on dust grains and PAHs also cause photoelectric ejection of electrons which then
collisionally heat the gas, a process called the grain photoelectric heating
mechanism (e.g., Watson 1972, Bakes \& Tielens 1994, Weingartner \& Draine 2001).
The gas cools via
emission in many IR lines. Emission from PDRs has been quite successfully
modelled (see Hollenbach  \& Tielens 1999 and references therein). The models
are usually parameterized by $G_0$, the ratio of the FUV flux to the Habing
(1968) FUV band flux of $1.6 \times 10^{-3}$ erg cm$^{-2}$s$^{-1}$
characteristic of the local interstellar radiation field, and $n$, the density
of the gas. For advecting (non-stationary) models, the flow velocity  of the
material through the PDR is also a parameter.
The FUV fields in PDRs near OB star-forming regions such as the Orion Bar or
M17 SW are typically very high, ($G_0$ $\sim$ $10^{4-5}$); the PDR
gas in these regions is inferred to be very dense with average densities $
\langle n \rangle
\sim 10^{4-5}$ cm$^{-3}$ (Tielens \& Hollenbach 1985b). On larger scales and 
in more evolved regions such as the
Rosette Nebula, the FUV  fields and the average densities
may be significantly lower by factors of $10^2-10^3$.

Infrared, sub-millimeter and millimeter wavelength observations of PDRs
indicate that gas in PDRs is inhomogeneous (van der Werf et al. 1993,
Luhman et al. 1998), which is not surprising since molecular clouds
themselves are observed to be highly inhomogeneous on a wide range of
scales from tens of parsecs down to  tenths of a parsec (e.g., Genzel 1991).
There is both indirect and direct evidence for clumpiness in PDRs.
Spatially extended fine structure line emission of neutral  and singly-ionized
carbon  near \ion{H}{2} regions shows that the FUV penetrates deeper
into the cloud than predicted by homogeneous models, suggesting that the gas
is clumpy in nature (Meixner et al. 1992, Stutzki et al. 1988, Steiman-Cameron
et al. 1997). Further, strengths of high excitation lines of molecular species
like CO J=14$\to$13 are observed to be greater than that predicted to arise from
gas with density inferred from other PDR species. It is thus inferred that 
 in regions such as  M17 SW and the Orion Bar, the PDR gas
consists of a ``low'' density (typically $ \sim 10^{3-5}$cm$^{-3}$)
component responsible for the extended CI and CII emission and high-density
clumps (n $\sim 10^{6-7} $ cm$^{-3}$)  which give rise to the high excitation
CO lines from their surfaces or, more precisely, at a depth of $A_v \sim $1
from their surfaces  (Burton et al. 1990, Meixner et al. 1992, Steiman-Cameron
et al. 1997, Hogerheijde et al. 1995). Clumps are also observed directly by
mapping PDRs at high spatial resolution in the   2 $\mu m$ lines of H$_2$ (e.g.,
Orion Bar, Luhman et al. 1998; Eagle Nebula, Allen et al. 1999). The small regions
of enhanced emission are interpreted as indicating the presence of high density
clumps ($n \sim 10^{6-7} $ cm$^{-3}$) with sizes $\le$ 0.02 pc, occupying a
very small volume filling factor $\sim 1-2 \% $. However, this interpretation is
not very conclusive, and in this paper we present evidence to show that small
clumps with such high densities may not exist in PDRs. 
 Small differences in temperature within the observed
regions can also explain the observed emission, as H$_2$ emission is highly
sensitive to gas temperature (Marconi et al. 1998).

Dense PDRs like the Orion Bar are spatially thin structures and if these PDRs are 
indeed clumpy, the clumps must have sizes smaller than the thickness of the warm PDR,  
$\sim 0.03$ pc for the Orion
Bar. The presence of small clumps affects the infrared spectrum from PDRs by
introducing a range of densities in the FUV-illuminated region and by introducing
enhanced advection in clumps that are photoevaporating. One of the goals of this
paper is to investigate the conditions under which small,
high-density  clumps exist in  PDRs.

Small clumps might not exist in significant numbers if the intense FUV field
photoevaporates them quickly. 
The incident FUV field heats up and
pressurizes the surface layers of the clumps, causing the layers to expand. This mass
loss may cause a complete photoevaporation of the dense clumps, and thus destroy them
on short timescales. 
 The mass loss timescales due to photoevaporation of an
FUV-exposed clump can be  estimated by assuming that the heated gas
at the surface flows outwards at its sound speed, $c_{PDR}$. The mass
density at the base of this heated flow, $\rho_{b}$, is lower than the mass density,
$\rho_c$, in the cold clump, as we will discuss below. (See Table 1 for a list
of frequently used symbols in this paper.)
The mass loss rate is given by $ dm_c/dt \sim 4 \pi r_c^2
\rho_{b} c_{PDR} $ and since $m_c \sim \rho_c r_c^{3}$, the photoevaporation
timescale is $t_{PE} \sim m_c/({dm_c/dt}) \sim r_c \rho_c/(c_{PDR}\rho_{b})$.
For a typical clump, and an FUV-heated surface at 1000K, this is approximately
$3 \times 10^3 (r_c/0.01 {\rm pc })  (\rho_c/\rho_{b}) $ years. Asymmetric
mass loss from the surfaces of very small clumps can rocket them
(e.g., Oort \& Spitzer 1955) out of the PDR back into the cloud.
Clumps that survive the rocket effect and remain in the
PDR are likely to lose significant fractions of their mass on
short timescales of order $\lesssim 10^{4-5} $ years, depending on
the ratio $\rho_c/\rho_b$. The clump evaporation timescale can be longer or 
shorter than its residence time in the
FUV-illuminated surface region of a GMC. From a frame of reference moving with
the advancing ionization front, interclump material and the clumps in the GMC
are advected into the PDR and ultimately flow into the \ion{H}{2} region. Large clumps
survive and they can penetrate the \ion{H}{2} region and affect its evolution and
structure, becoming Evaporating Gaseous Globules or ``EGGs'' (Bertoldi \& McKee
1990).

In the present paper, we focus on the heating of an individual clump in a PDR
by FUV photons from  a nearby O or B star. We study a range of FUV fluxes
incident on a range of clump sizes and determine the photoevaporative lifetimes
of the clumps and the evolution of their structure. Our investigation is
analogous to the study by Bertoldi (1989) and Bertoldi and McKee (1990) on the
effects of EUV heating of clumps in \ion{H}{2} regions, and a generalization of
the studies by Johnstone et al. (1998) and St\"{o}rzer  \& Hollenbach (1999) of the
evaporation of small clumps and protoplanetary disks (``proplyds'', O'Dell et
al. 1993) by EUV and FUV photons.

In a subsequent paper (Gorti \& Hollenbach, in preparation) we will discuss the
cumulative effect of O and B stars on Giant Molecular Clouds (GMCs), and the
relative importance of EUV and FUV photons in dissociating and destroying GMCs.
Ultimately,  we are interested in the role of FUV radiation in regulating
star-formation in GMCs. Destruction through photodissociation and
photoevaporation of potentially star-forming clumps/cores in GMCs reduces the
star-forming efficiency of GMCs, and the FUV field may thereby play an
important role in regulating or limiting star formation. Photoevaporation of
the outer envelope of a collapsing isothermal sphere by FUV radiation could
rapidly deplete the material in the envelope, and may limit the final mass of
the star formed. On the other hand, compression of non-collapsing clumps by shock
waves driven by the warm surface gas could possibly drive the inner cores to
instability and gravitational collapse, triggering star formation.

In \S 2 of this paper, we discuss the present understanding of GMC structure 
and existing models.  In \S 3, we describe the evolution of an \ion{H}{2} 
region and its surrounding PDR as they propagate into the GMC. Timescales relevant
for the FUV heating  of a clump in various contexts are briefly discussed
in \S 4. In \S 5 we present a simple analytical model which provides an
understanding of the underlying physics of FUV-induced clump evolution. We
consider various possible evolutionary scenarios and their dependences on clump
parameters and the strength of the FUV field. For the more general case, a 1-D
numerical hydrodynamics code has been developed and these results are described
(\S 6). In \S 7, we estimate the significance of the rocket effect due to
asymmetric mass loss and its effect on clump lifetimes in the PDR, and in
\S 8 we discuss the implications of our results for observations of PDRs in
star-forming regions. Application to observed PDRs is made in \S 9.  We conclude
with a summary of the present investigation (\S 10)

\clearpage
\begin{table}[f]
\caption{List of symbols frequently used in this paper}
\begin{tabular}{cl}
& \\
\hline
Symbol & Meaning \\
\hline
$c_c$ & Thermal sound speed in cold clump gas \\
$c_{PDR}$ & Sound speed of heated, warm PDR gas \\
$G_0$ & Strength of the FUV field in Habing units, $1.6 \times 10^{-3}$ erg s$^{-1}$
         cm$^{-2}$ \\
$N_0$ & Fiducial column density equal to $2 \times 10^{21}$ cm$^{-2}$\\
$n_{c0}$ & Initial gas number density of clump \\
$m_{c0}$ & Initial mass of clump \\
$r_{c0}$ & Initial radius of clump \\
$t_c$ & Sound crossing time in cold clump gas, $r_{c0}/c_c$ \\
$t_e$ & Expansion timescale for clumps with $\eta_{c0} > \eta_{crit}$\\
$t_{FUV}$ & Timescale on which the cold clump gas gets heated to the maximum temperature\\
 &         in a given  FUV field \\ 
$t_{PE}$ & Photoevaporation timescale \\
$t_s$  & Shock-compression timescale for clumps with  $\eta_{c0} < \eta_{crit}$ \\
$v_{IF}$ & Velocity of the Ionization Front moving into the GMC\\
$X_{PDR}$ & Thickness of the PDR in the interclump medium of the GMC, bounded by the IF and DF \\
$\alpha$ & Initial ratio of turbulent to thermal pressures in the cold clump gas\\
$\beta$ & Initial ratio of magnetic to thermal pressures in the cold clump gas\\
$\gamma$ & Power law exponent of the variation of density with magnetic pressure \\
$\delta_0$ & Initial thickness of PDR shell on the surface of a clump ($=N_0/n_{c0}$) \\
$\eta_{cr}$ & Critical column density for complete photoevaporation of 
pressure-confined clumps\\
$\eta_{c0}$ &  Normalized initial column density to centre of clump, $n_{c0} r_{c0}/N_0$\\
$\eta_{crit}$ & Critical column density in a given FUV field
  towards which unconfined clumps evolve \\
$\eta_f$ & Final column density of pressure-confined clumps in an FUV field \\
$\lambda$ & Photoevaporation parameter, $(1+\alpha+\beta)(\eta_{c0}-1)/
(2(2\nu^2+\alpha))$ ; $\lambda=1$ for $\eta=\eta_{crit}$ \\
$\nu$ & Ratio of sound speeds in warm and cold gas, $c_{PDR}/c_c$ \\
\hline
\end{tabular}
\end{table}
\clearpage

%
\section{Clumps in Giant Molecular Clouds}
Giant Molecular Clouds are observed to have a hierarchical
substructure that is highly inhomogeneous, with density enhancements such as
cores, clumps and filaments on all observable scales, from $\sim 100$ pc down to
$\sim 0.1$ pc. We use the term `` massive cores'' to denote  large scale
density enhancements ($\ga 1 $ pc in size) and ``clumps'' to denote  smaller
scale structures ($\sim$ few tenths of pc or less in size)\footnote{Note 
that our terminology of cloud substructure is the reverse
of what is sometimes used in the literature. We call the large scale structures
`` massive cores'' and the small scale structures ``clumps'' to be consistent with
observational work on dense regions in PDRs, wherein the density enhancements
are usually called ``clumps''. Our notation is also consistent with the notion
of ``hot cores'', where massive stars form.}.
GMCs and their more massive constituent cores are
strongly self-gravitating, and are supported against collapse by turbulence and
magnetic fields. The internal velocity dispersions of the
substructures in molecular clouds are found to scale with their sizes, with decreasing
non-thermal support on the smallest scales (Goodman et al. 1998). The column
density through a GMC or a typical
GMC substructure is found to be approximately a constant
$N \sim 10^{22}$ cm$^{-2}$, although significant variations
($N \sim 10^{21-23}$ cm$^{-2}$)
about this typical column may occur. (For a recent review on GMCs, see McKee 1999).

The smaller substructures or clumps are not all gravitationally bound and the
exact physical nature of these clumps remains uncertain (Williams et al.
2000). They have been interpreted as temporary fluctuations in density caused
by supersonic turbulence within the cloud (e.g., Falgarone \& Phillips 1990,
Scalo 1990). This view is supported by their apparent fractal nature, which
seems to show the same structure on smaller scales down to a few tenths of a
parsec (Elmegreen 1999), and their observed supersonic linewidths (Blitz \&
Stark 1986). In this picture, the gravitationally unbound clumps are transient
objects formed by converging supersonic turbulent flows. The non-gravitating
clumps are therefore being constantly formed and destroyed, and have short
lifetimes of the order of their sound crossing time,  $t_c \approx r_{c0}/c_c$,
where $c_c$ is the clump sound speed, and $r_{c0}  $ is the initial clump radius.
The effect of the sudden turn-on of an FUV field on such clumps
is to drive a shock into the clump, producing a smaller, denser clump with a
warm  evaporative outflow. Surprisingly, the evaporative timescale is
now  somewhat longer than  $t_c$, as we will show. Assuming that the FUV flux
does not strongly affect the turbulence so that the formation timescale remains the 
same, the presence of an FUV field will not greatly change the non-gravitationally
bound clump abundance, but it will compress them  and alter their
thermal structure and dynamical evolution.

Another interpretation is that small bound clumps as well as small unbound clumps
represent stable physical entities confined by interclump pressure.
Observations through molecular line spatial
maps and velocity channel maps or position-velocity diagrams indicate the
presence of dense coherent structures. Indirect evidence suggests the presence of
a low density ($\sim 1-2$ orders of magnitude less than the clumps) ``interclump
medium'' (ICM) (see Williams et al. 1995 and references therein). The ICM
exhibits high velocity motions ($\Delta v \approx 10$ km s$^{-1}$)
and therefore may have significant pressure,
which could be of thermal, magnetic or turbulent origin.
Clumps in this picture are  confined by the ICM, and some of them may be
self-gravitating (Maloney 1988,  Bertoldi \& McKee 1992). Pressure-confined
clumps live much longer and presumably form more slowly than the gravitationally
unbound clumps of the turbulent model. Therefore, FUV heating of these
clumps and their destruction through photoevaporation will have a much more pronounced
effect on lowering the clump abundance in this interpretation.

%
\section{The Overall Evolution Of A Massive Star Forming Region} Massive stars
form in the central regions of clumpy, massive,  cores ($M \gtrsim 1000 M_\odot$), 
of radius $\sim 1 $ pc inside GMCs (see review by Evans 1999).
These massive  cores contain clumps whose densities may reach $n_c \sim 10^{5-7}$
cm$^{-3}$, or even higher (Plume et al 1997), and the clumps are surrounded
by interclump gas with lower, but uncertain densities, possibly  $n_{ICM}
\sim 10^{3-5}$ cm$^{-3}$.
When a massive star forms, the EUV photons from the star immediately ionize the
nearby interclump gas, and form an \ion{H}{2} region. The clumps in the \ion{H}
{2} region are also exposed to the ionizing radiation and  evaporate (Bertoldi
1989, Bertoldi and McKee 1990).  The FUV photons penetrate further through the
interclump gas  beyond the interclump ionization front (IF) and dissociate the
molecular gas, forming a dissociation front (DF) and a PDR. 

In the earliest stages of evolution when EUV photons  emerge from the 
forming O/B star, the IF moves as an R-type front. 
However, the IF quickly stalls and becomes D-type, at which point the high
pressure \ion{H}{2} region drives a $\sim $ 10 km s$^{-1}$ shock into the neutral
interclump gas beyond the IF. It is instructive to obtain a sense of sizescale
at the R/D type transition. Assuming the massive star emits $\phi_i$ Lyman
continuum photons per second, the Str\"{o}mgren radius at this juncture is $r_s
\approx 0.1 \phi_{49}^{1/3} n_{e4}^{-2/3} $pc, where $\phi_{49} = \phi_i/10^{49}
$s$^{-1}$ and the electron number density $n_{e4}=n_e/10^4$ cm$^{-3}$. If the
central star has an FUV luminosity of $L_{FUV} = 10^5 L_{\odot}$
(typical of stars with $\phi_{49} \approx 1$), the FUV flux
incident on the surrounding PDR is characterized by $G_0 \approx 10^5$. The
thickness of the PDR is approximately $10^{22}$cm$^{-2}/n_{ICM}$ or $0.3
n_{ICM4}^{-1}$ pc. The DF initially advances ahead of the shock, but as
photodissociation approaches equilibrium with H$_2$ formation, the DF slows down.
Eventually, the shock moves ahead of the DF, and the entire PDR is contained
in the post-shock shell. The shock  now impacts cold molecular interclump
gas. Hill  \& Hollenbach (1978) showed that at this stage the shock velocity
has slowed to $\lesssim 3 $km s$^{-1}$, and that, if $n_{ICM} \gtrsim 10^4$ cm$^{-3}$,
the shock completely stalls in $t \la 10^5$ years. Therefore, there is
an initial stage ($t \lesssim 10^5 $ years, for dense ICM) in the evolution
of the clumpy PDRs surrounding \ion{H}{2} regions where the clumps are both
shocked and exposed to the photoevaporating effects of the FUV radiation. This
can be followed by a stage where the shock has stalled and unshocked clumps move 
into the FUV-illuminated zone.

Two configurations are possible after the shock stalls. If the \ion{H}{2} region
is still completely embedded, it becomes stationary with its pressure matched by
the molecular cloud pressure and the ionizing photons completely absorbed by
recombining electron/proton pairs in the \ion{H}{2} region. Much more likely is 
the situation where the \ion{H}{2} region breaks through the surface of the cloud
and becomes a ``blister'' \ion{H}{2} region. In this case, the ionized gas can
expand away from the cloud,  and into the interstellar medium. The IF then
slowly eats its way into the PDR and the DF similarly advances into the
molecular cloud. The velocity of the IF advancing into the PDR is of order
$v_{IF}  \sim 1 $ km s$^{-1}$, just enough so that the flux of particles through
the IF can balance the heavily attenuated flux of EUV  photons reaching the IF
(see Whitworth 1979, Bertoldi \& Draine 1996, St\"{o}rzer \& Hollenbach 1998).
In such a case, the clumps entering the PDR may not have been previously shocked
and compressed. Further, in the turbulent GMC model, clumps form and dissipate
continually and those formed in the PDR after the passage of the shock are not
affected by it. In both the PDR around the Trapezium stars in Orion and in the
M17 SW PDR (Meixner et al. 1992), there is no evidence for velocity shifts between
PDR gas and the ambient cloud gas, or, in other words, no evidence for a shock
at the present time.  In this paper, we focus only on the photoevaporation
of clumps, whether or not they have been shocked, and do not follow the potential
shredding or flattening of a clump by the passage of a shock wave (see, e.g.
Klein et al. 1994).

The thickness or column depth of a PDR has been defined by Tielens \& Hollenbach
(1985) to include all the gas and dust where FUV photons play a significant role
in the gas heating or chemistry. By this definition, in regions like Orion or
M17, the PDR extends to columns of $\sim 10^{22}$ cm$^{-2}$ and includes gas which
is cool and molecular (H$_2$, CO), but where FUV still photodissociates
O$_2$ and H$_2$O. We are interested in the surface column N$_0$ of
the PDR which is heated to the highest temperatures (T$_{max}\sim 100-3000$K,
depending on $G_0$ and $n$). For G$_0/n > 4 \times 10^{-2}$ cm$^3$, N$_0
\approx 1-3 \times 10^{21}$ cm$^{-2}$ (equivalent to a dust FUV optical depth
of order unity) and is nearly equal to the column where atomic hydrogen converts
to H$_2$ in stationary PDRs (Tielens \& Hollenbach 1985).  We shall call
this region the ``warm PDR'' hereafter to distinguish it from the entire PDR.
The thickness of the warm PDR, N$_0/n_{ICM}$, is of order $0.03-0.1 n_{ICM4}^{-1}$
pc. The crossing timescale for the IF through the warm PDR is
$t_{PDR}\approx   0.3-1 \times  10^5 n_{ICM4}^{-1} v_{f5}^{-1}$ years,
where $v_{f5}=v_{IF}/(1$ km s$^{-1}$). Gas and dust (and clumps)
from the opaque molecular cloud
interior are therefore advected on these timescales from the shielded cloud,
through the warm PDR, and across the IF into the \ion{H}{2} region. Figure 1
shows a schematic diagram of FUV radiation from nearby O/B stars impinging
on massive cores (large sizescales, e.g., Rosette) and clumps (small scales, e.g.,
Orion).

\clearpage
\begin{figure}[f]
\includegraphics[bb=50 250 550 600]{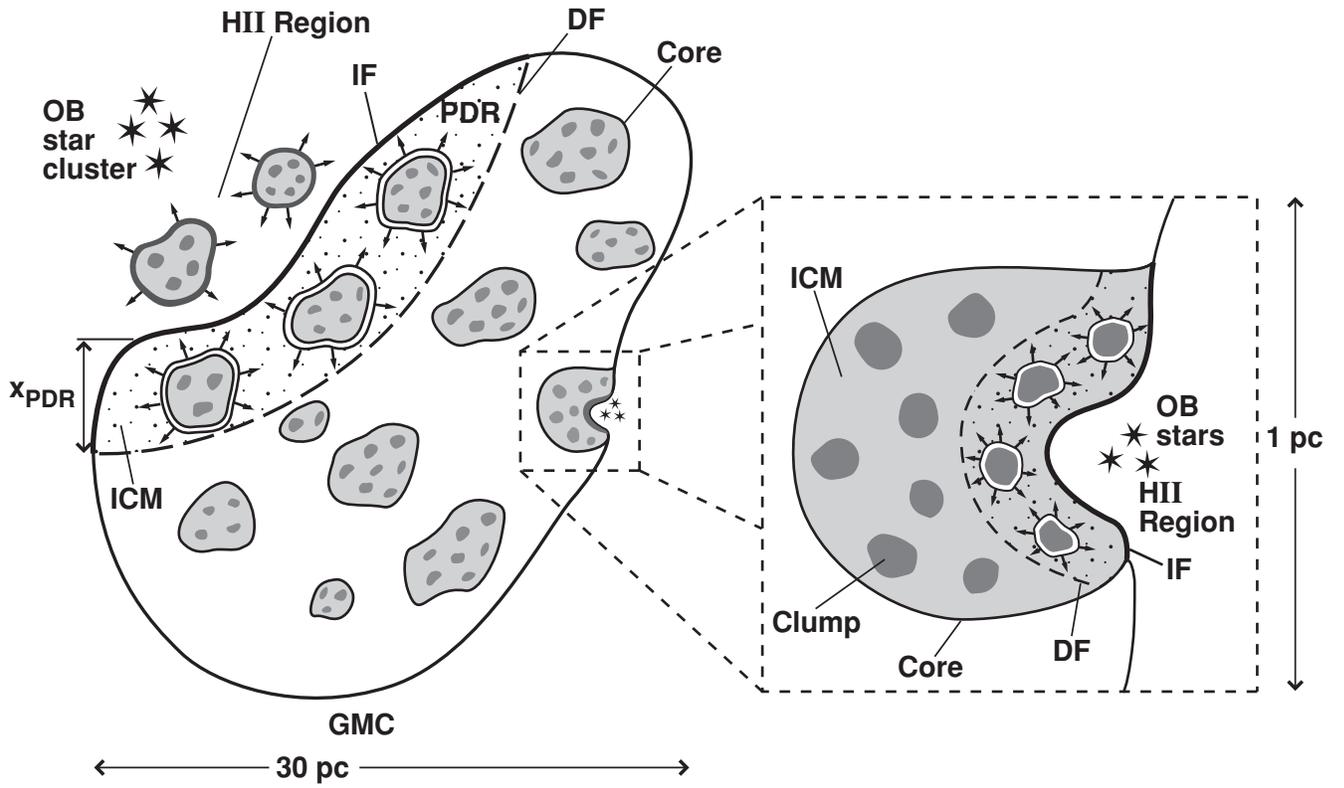}
\caption{A schematic diagram of PDRs on large scales (e.g. Rosette) and small
scales (e.g. Orion Bar). The figure shows blister \ion{H}{2} regions formed
by the massive stars, and the PDR bounded by the ionization and
dissociation fronts. Clumps in the PDR are exposed to FUV radiation and
lose mass by photodissociation and/or photoevaporation.}
\end{figure}

\clearpage
%
\section{Impulsive heating versus slow heating}
FUV-illuminated clumps are heated at their surfaces on timescales $t_{FUV}$
that could be either fast or slow compared to the internal sound crossing
timescale within the clump, $t_c = r_{c0}/c_c$. Impulsive heating implies that
the FUV heating of the clump surface layer, characterized by column $N_0$, occurs
with $t_{FUV} \ll t_c  $. In this case, the outer region is heated more
quickly than the entire clump can respond, and attains a much higher pressure
than the cold central region. A shock propagates inward and a heated
photoevaporative flow expands off the surface.

Several situations in clumpy molecular clouds may lead to $t_{FUV}  \ll t_c $.
For a clump with $r_{c0} \sim 10^{17}$ cm and $c_c \sim 0.3$ km s$^{-1}$ (T $\sim$
10K), $t_c \sim 10^5$ years. The OB star may ``turn-on'' on a timescale $< 10^5$
years. Even after this initial flash of FUV flux, the clumps may shadow each
other so that $t_{FUV}$ is given by the time for a clump to move out of shadow,
which is of order $ r_{c0}/v_c$, where $v_c$ is the clump velocity. In a
turbulent, supersonic medium $v_c$ may be larger than $c_c$ so that $t_{FUV}
 <  t_c$. Similarly, if clumps are formed in turbulent gas
by supersonic converging flows, the FUV turn-on timescale is roughly the time
for the clump to form, which again is $t_{FUV}\approx r_{c0}/v_t$, where $v_t$
is the supersonic turbulent speed. Therefore, impulsive heating is  likely
the most appropriate approximation for the turbulent model of clumps in
molecular gas.

Let us consider the other situation of long-lived clumps confined by the ICM.
Here, slow heating is generally the best approximation, but in certain
conditions, impulsive heating may be more appropriate. 
A common situation after the turn-on of the OB star may be the advancement of
the ionization front and PDR front into the opaque molecular cloud, which occurs
at speeds $v_{IF} \sim 1$ km s$^{-1}$. In this situation $t_{FUV}$ is the time for the
DF and the IF  to move so that the clump is
one FUV optical depth closer to the IF, or $t_{FUV}
\approx t_{PDR} = X_{PDR}/v_{IF}$, where $X_{PDR} = N_0/n_{ICM}$ is the thickness of
the heated PDR zone in the ICM of density $n_{ICM}$. The criterion for
impulsive heating $t_{FUV} <  t_c$, is more easily met by larger clumps that
have longer sound crossing timescales, as $t_{FUV}$ is the same for all clumps.
This gives us a limiting size for clumps likely to be impulsively heated,
\begin{equation}
r_{c0} >X_{PDR} {{c_c}\over{v_{IF}}} \approx (0.3-1)X_{PDR},  \label{b}
\end{equation}
which is of the order of the thickness of the interclump PDR region itself.
Clumps with radii smaller than this are slowly heated as they emerge from the
opaque molecular cloud into the FUV-irradiated regions. Since we only consider
clumps {\em within} PDRs that have $r_{c0} < X_{PDR}$,  pressure-confined clumps
are generally slowly heated in our analysis.  

Clumps that undergo slow heating, $t_{FUV} \gg t_c $, adjust quasi-statically
to the changing FUV flux. Long-lived, small, pressure-confined
clumps advecting into FUV-irradiated regions provide a prime example of this
slow evolution. Such clumps develop a warm PDR surface
which slowly gets warmer, thicker and less dense. The clump and the heated PDR
surface evolve in near pressure equilibrium, with the density of the surface
decreasing with the rise in its temperature, such that $n_{PDR} T_{PDR} = n_c
T_c = $ pressure of the ICM (assuming thermal pressure dominates on these small
scales). At any given time, the evolutionary state of a slowly heated clump is
the same as the {\em final} state of an impulsively heated clump exposed to
the same local FUV flux.  However, a gradually heated clump constantly adjusts
its PDR surface to maintain pressure equilibrium as the PDR slowly heats. In
contrast, an impulsively heated clump undergoes shocks and supersonic expanding
flows in its attempt to attain equilibrium. (We note that there is no final 
equilibrium state if the interclump pressure is zero). Rapid photoevaporative mass
loss can cause an acceleration due to a rocket effect in very small clumps and
move them back into the shielded cloud. Pressure-confined small clumps, on the
other hand, are not rocketed as they evolve quasi-statically. In a given FUV field
and at a given ICM pressure, the final equilibrium state of a small clump
is either a completely photodissociated, warm $(T_{PDR})$, and expanded 
(but with $P=P_{ICM}$) region;
or it may consist of  a shielded cold molecular core with the same temperature,
density and pressure as the original clump, but with a heated and more diffuse
protective surface layer of PDR material. In \S 5.3 we provide analytic
solutions for these equilibrium structures. However, in \S 5 and \S 7 we primarily
focus on the interesting physical processes which occur for clumps that are
impulsively heated.
%
\section{Analytical models for FUV photoevaporating clumps}
We first  provide approximate analytic solutions for the time evolution of a
clump suddenly exposed to  FUV radiation, using a few simplifying assumptions.
As described in  \S 2, the nature of the clumps themselves is not well
understood, and we consider two simple analytic models that qualitatively correspond
to the two main interpretations. Turbulent unbound clumps are modelled as
constant density structures in vacuum. A clump in vacuum disperses on a sound
crossing timescale, in the absence of FUV illumination. Pressure-confined clumps
are also treated as being of constant density, but have a surrounding ICM with
equal pressure. In our models, the ICM only serves to confine the clumps and
does not itself evolve or otherwise affect the clump evolution. We seek solutions
for the time evolution of the mass, size and mass loss rate of a photoevaporating
clump.

\subsection{Initial conditions and assumptions}
Clumps are assumed to be  dense, small  spheres of gas, supported by thermal,
turbulent and magnetic pressures.  We assume  that the magnetic field $B$
scales with a constant power of the density, so that the magnetic pressure is
\begin{equation}
\label{gamma} P_B = B^2/8\pi \propto n^{\gamma}.
\end{equation}
The initial ratio of turbulent and magnetic pressures, $P_{NT}$ and $P_B$, are
set by two dimensionless parameters,
\begin{equation}
\alpha = P_{NT}/ P_T; {\rm \hspace{1cm}} \beta=P_B/P_T
\end{equation}
where $P_T$ is the initial thermal gas pressure in the clump.
A clump may have significant turbulent support and observations of clumps
in star-forming regions (Jijina et al. 1999) indicate values of $\alpha$
ranging from $\sim 0$ in cold, dark clouds to $\ga 2$ in regions of massive
star formation. There is observational evidence to suggest that turbulent
support in clumps decreases on the smaller scales of star-forming clumps
($r_{c0} \la 0.1$ pc), where
$\alpha \la 1$ (Goodman et al. 1998). Measurements of magnetic fields and
hence $\beta$ and $\gamma$ are difficult, but present 
observational data suggest a wide range of values for $\beta$, from $\sim 0$
to a few (Crutcher 1999). The value of $\gamma$ ($\sim 1-2$) describes how the
B field responds to a relatively sudden change in gas density.  For a frozen,
uniform magnetic field threading a spherical clump, $B\propto n^{2/3}$ or
$\gamma=4/3$. For our standard case we adopt values of $\alpha=1, \beta=1$, 
and $ \gamma=4/3$. We later discuss the sensitivity of our results to these 
choices for $\alpha$, $\beta$ and $\gamma$.
As discussed in \S 4, for clumps confined by an ICM and advecting into
FUV-illuminated interclump zones, the assumption of
$t_{FUV}/t_c < 1$ is equivalent to $r_{c0} \gtrsim (0.3-1) X_{PDR} $. Therefore,
in this case we treat only clumps above a certain minimum size, which is of order
the size of the thickness of the heated PDR of the ICM.  We do not explicitly
include self-gravity in our analytical model, and only qualitatively estimate the
effects of gravity on clump evolution. We treat gravity more quantitatively in the
numerical models.

The FUV radiation field as seen by the clump is spherically asymmetric, and
stronger at the surface facing the source. For simplicity, however,
we assume a spherically symmetric FUV field when calculating the internal dynamical
evolution and the thermal structure of the evaporating clump.
Typical albedos of interstellar dust are $\sim 0.5$ and some of this scattered
radiation is directed backwards (dependent on the phase factor $g$, see
Henyey \& Greenstein 1941), resulting in a backscattered FUV flux on the
shielded clump surface about $0.1-0.2$ times that on the directly illuminated
surface (e.g., D\'{e}sert et al. 1990, Hurwitz et al. 1991). 
Although the posterior intensity is diminished compared
to that on the front surface, the PDR surface sound speed is not a
sensitive function of the incident FUV flux $G_0$ (Kaufman et al. 1999). For
field strengths $G_0 \sim 10^{3-5}  $ near typical \ion{H}{2} regions and densities
$n \gtrsim 10^{3-6}$ cm$^{-3}$ characteristic of clumps there, even a back flux
10\% of that on the clump surface facing the source would heat the gas to nearly 
the same sound speed ($c_{PDR} = 3$ km s$^{-1}$ on the source side, $c_{PDR} = 
2$ km s$^{-1}$ on the shadowed side). Therefore, the assumption of 
spherical symmetry is as good as ignoring the flux of FUV photons on the
backside when calculating the thermal structure and internal dynamics. However,
the asymmetry of the flows from the front and back surface will cause a
rocket effect on the clumps and induce motion away from the source, as first
recognized by Oort \& Spitzer (1955). We discuss momentum transfer due to
mass loss of the clumps in \S 7, and estimate sizes of clumps that gain velocities
high enough to move out of the PDR and back into the molecular cloud.

In our analytical models (but not our numerical models) we assume that the
transition from the warm outer PDR layer to the cold molecular core of a clump
is very thin, so that the clump can be modelled as an isothermal core with sound
speed $c_c$, surrounded by a PDR envelope of sound speed $c_{PDR}$.
The gas in the core always remains isothermal in our analysis, even when
shock-compressed to high densities.  For our
pressure-confined clumps, it is assumed that the ICM is of low density and does
not absorb any FUV photons. In order to obtain analytic solutions, we do not
fully treat transient phenomena and emphasise quasi-steady state aspects of the flow.
\subsection{Analytic model for unconfined clumps}
We begin by considering the evolution of a simple configuration: a cloud
consisting of clumps that are spheres of radius $r_{c0}$ and constant initial
density $n_{c0}$, immersed in a vacuum. We refer to such clumps hereafter as
``turbulent clumps''. These are dynamic structures that in the absence of
FUV heating expand at roughly their sound speed, $c_c$. When such a clump is exposed
to an FUV field, the photon flux is attenuated by dust and instantaneously
heats an outer column
of N$_0 \sim 2 \times 10^{21} $ cm$^{-2}$ (thickness $\delta_0 = N_0/n_{c0}$) to
relatively high temperatures and sound speeds (for example, near \ion{H}{2} regions
with $G_0 \sim 10^{4-5}$ and $n_{c0} \sim 10^5$ cm$^{-3}$, $ T \sim 1000$K and
$ c_{PDR}\sim 3$ km s$^{-1}$).

The evolution of the clump is determined by its initial radial column density
from the centre outwards, $n_{c0}r_{c0}$, and the ratio $\nu$ of sound 
speeds in the FUV-heated region and the cold clump material,
\begin{equation}
\label{nudef}
\nu=c_{PDR}/c_c
\end{equation}
The parameter $\nu$ can be thought of as a measure of the strength of the FUV
field incident upon the cold clump. We introduce the dimensionless parameter $\eta_{c0}$,
\begin{equation}
\label{etadef}
\eta_{c0} = {n_{c0} r_{c0}}/{N_0} = r_{c0}/\delta_0 ,
\end{equation}
for the initial column density to the centre of  the clump. Note that
$\eta_{c0}$ is a measure of the mean column density through a clump.

For turbulent clumps with $\eta_{c0} \le 1$,  the initial clump column density
is less than $N_0$, and the entire clump is immediately heated and
photodissociated by the FUV flux. Effectively, therefore, the FUV flux
accelerates the expansion of the cold clump by heating it throughout, decreasing
the expansion timescale by a factor given by the ratio of sound speeds at the
two temperatures, $c_c/c_{PDR}$, or $1/\nu$.
For clumps with $\eta_{c0} > 1$, the FUV flux heats an outer shell of gas to 
a higher temperature, thereby increasing its pressure. If the pressure of 
the outer PDR shell is sufficiently high, it drives a shock to the centre of 
the clump rapidly compressing it, and the compressed clump proceeds to 
evaporate on a somewhat longer timescale than the expansion timescale 
in the absence of an FUV field. 
On the other hand, if the pressure in the heated outer PDR shell is low, 
the shock does not make it to the centre of the clump. The shock dissipates, 
followed by an expansion of the clump, until finally the photoevaporative 
flow halts the expansion and proceeds to shrink and compress the gas. There 
are thus two distinct evolutionary scenarios for photoevaporating turbulent 
clumps with $\eta_{c0} > 1$.

For brevity in the main text, the details of our analysis and the derivations of
the results are presented in the appendices to the paper. We now discuss our
solutions and describe the physics of FUV-heated clump evolution and refer 
the interested reader to the appendices for the full analysis.

Clumps exposed to a given FUV field, measured by the parameter $\nu$, are either
compressed by shocks or expand to adjust their column density to a critical value,
$\eta_{crit}$. This critical column density, $\eta_{crit}$ can be derived
from conservation of mass and from the condition of 
pressure equilibrium at the surface of the cold clump gas (see 
Appendix A for details), and is defined by the relation
\begin{equation}
{{2(2\nu^2+\alpha)}\over{\eta_{crit} -1}} +  \beta \left(
{{2}\over{\eta_{crit}-1}}
\right)^{\gamma}= 1 + \alpha + \beta \label{aetext}
\end{equation}
Recall that $\alpha$ and $\beta$ are the initial ratios of turbulent to thermal
pressure and magnetic to thermal pressure, respectively, and that the magnetic
pressure scales as $n^{\gamma}$. For typical values of these parameters, the 
second term on the LHS is negligible and equation~(\ref{aetext})
can be used to define a photoevaporation parameter, $\lambda$, as
\begin{equation}
\lambda = {{(1+\alpha+\beta)(\eta_{c0}-1)}\over{2(2\nu^2+\alpha)}}. 
\label{defltext}
\end{equation}
For clumps with initial column densities $\eta_{c0}=\eta_{crit}$, the
photoevaporation parameter $\lambda=1$. Figure 2 shows an $\eta-\nu$ 
parameter plot with equation~(\ref{aetext})  depicted for the standard case 
with $\alpha=1, \beta=1$, and $\gamma=4/3$, and for two possible extremes of 
these parameters. The result is not very sensitive to the choice of the 
parameter $\gamma$, and $\eta_{crit}$ only depends on the sum of $\alpha$ and $\beta$.
For the standard case and with $\eta_{c0}\gg 1  $ and $\nu \gg 1  $
which is typical of many PDRs, equation~(\ref{aetext}) can be approximated
as $\eta_{crit}  \approx 4\nu^2/3$. The evolution of FUV-heated clumps is 
determined by where they initially lie in the $\eta-\nu$ parameter space.

\clearpage
\begin{figure}[f]
\includegraphics[scale=0.4,bb= 20 145 580 710]{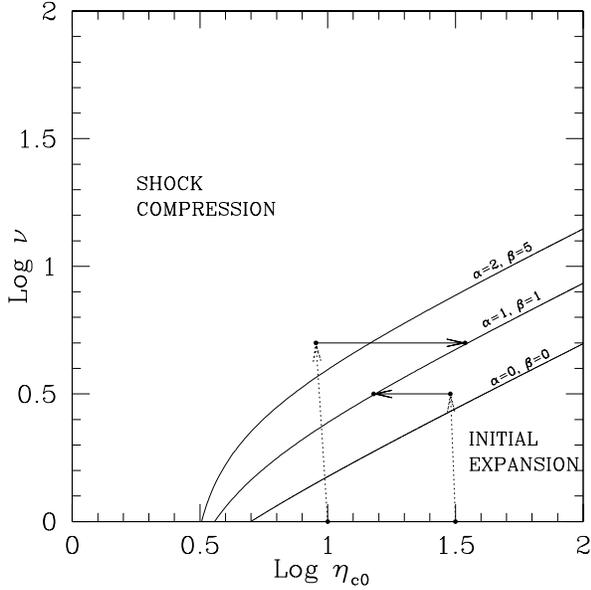}
\caption{Logarithmic parameter plot of the ratio $\eta_{c0}$ of the
initial column density through the clump  to the fiducial column $N_0$,
and the ratio $\nu$ of the sound speed in the heated PDR gas to that
in the cold clump gas, for impulsively heated clumps. The critical column
density  $\eta_{crit}$ is plotted against  $\nu$
for different values of $\alpha$ and $\beta$, with
$\gamma=4/3$. Also shown are representative trajectories of 
column density evolution for clumps
with initial column densities less than, and greater than, $\eta_{crit}$.
Note that, in this case, the abscissa refers to the instantaneous column density
through the clump. At $t=0$, $\nu=1$ and the clump has a column $\eta_{c0}$.
With the turn-on of the FUV field, $\nu$ increases rapidly.
A clump with $\eta_{c0} < \eta_{crit}$ is then compressed by
a shock to $\eta_{crit}$ and moves right on this parameter plot and then
proceeds to photoevaporate steadily. A clump with $\eta_{c0} > \eta_{crit}$ 
expands so that its column density becomes equal to $\eta_{crit}$ and  
continues to photoevaporate. The initial turn-on of the FUV field (dotted line)
moves the column slightly to the left because a column $N_0$ is quickly dissociated
and lost from the clump. Gravity is ignored in this figure (see Figure 5).}
\end{figure}
\clearpage

Clumps with an initial column density
$\eta_{c0} < \eta_{crit}$ evolve through a shock-compression phase, until
their dimensionless column density increases to  $\eta_{crit}$ (see Appendix B).
 Clumps with
higher initial column densities expand until their column density decreases to
$\eta_{crit}$ (see Appendix C). The evolution of clumps in two representative cases is 
qualitatively shown in Figure~2
through their trajectories in $\eta-\nu$ space. Once the clump column 
density attains its critical value $\eta_{crit}$, further evolution is 
similar in both cases, and the clump column density at any later time 
remains constant at this critical value.
We note that this would imply that observed clumps in a given PDR should typically 
have the same column density, the critical value $\eta_{crit}$ given by the 
ambient FUV field. This is valid for clumps photoevaporating into a vacuum and
only violated in the transient early stages which are typically very short, of
order $t \approx r_{c0}/c_{PDR}$. From equation~(\ref{aetext}), clumps at 10K heated
on their surfaces to 100 K
(or $\nu \sim 3$  ) should have columns $\eta_{crit} N_0  \sim 3 \times
10^{22}$ cm$^{-2}$, and for $T \sim  1000K$ or $\nu \sim 10, \eta_{crit} N_0
\sim 3 \times 10^{23}$ cm$^{-2}$. As photoevaporation
proceeds, and clumps lose mass, they shrink in size and their average density
increases, so that the column through the clump remains the same, $n_c(t) r_c(t)
= \eta_{crit} N_0$.

\paragraph{Shock compression}  In this case, the initial column
density is larger than $N_0$, but smaller than $\eta_{crit} N_0$.
The FUV can only penetrate through an outer shell of thickness
$\delta_0 = N_0/n_{c0} < r_{c0} $. This shell is heated to a higher
temperature and the corresponding increase in pressure causes the outer
shell to expand at $c_{PDR}$  radially outwards at the edge
and inwards into the clump at the inner
boundary. Figure 3
shows a schematic diagram of the evolution of a shock-compressed clump.

\clearpage
\begin{figure}[f]
\includegraphics[scale=1,bb=100 230 500 610]{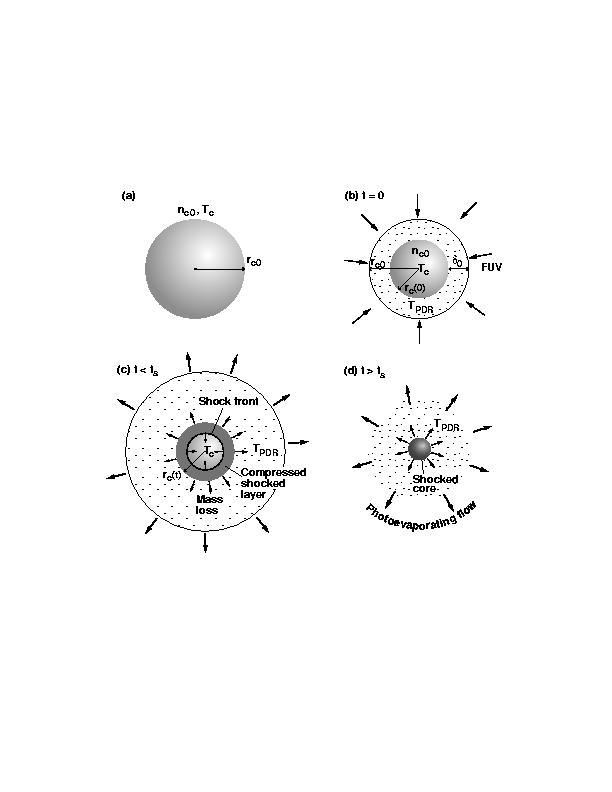}
\caption{Schematic diagram of  evolution of shock-compressed clump. The cold clump
is indicated by the lightly shaded region, the dotted region represents the
heated warm PDR surface of the clump, and the dark shaded region shows shock-compressed
gas. Panel c shows the propagation of the shock
front into the clump and panel d depicts the evolution of the clump after being
compressed by the shock into a small dense core.}
\end{figure}
\clearpage

As  the shell expands, the column density $N_0$ through it is
maintained by further penetration of FUV photons, and heating of
more cold clump material. The
pressure at the base of the PDR shell is initially higher than the cold
clump gas. This pressure difference drives a shock wave into the clump,
which rapidly propagates to the centre. Behind the shock the cold clump
gas is compressed to a pressure approximately equal to the pressure at the
base of the PDR shell.  The shock travels inward at roughly $c_{PDR}$
and the entire clump gets compressed in a time $t_s \simeq  r_c(0)/c_{PDR}$,
where $r_c(0)=r_{c0} - \delta_0$ is the initial radius of the cold
clump when the FUV flux is turned on (Fig. 3b). The shock reaches the centre and
compresses the clump to a radius $r_c(t_s)=r_s$.
The shock-compressed gas forms a high-density core, such that the
pressures in the core and at the base of the PDR flow are approximately equal. Further
mass loss from the clump is from the surface of this core. The time
evolution of the mass and radius of the clump for $t > t_s$ are given by
the following equations for our standard case, where $\alpha=1$, $\beta=1$,
 and $\gamma=4/3$. For the general solution, refer to Appendix B.
\begin{equation}
r_c(t>t_s) = \left(\left({{r_s}\over{r_c(0)}}\right)^{5/4} - {{10\eta_{c0} \nu}
\over{3(\eta_{c0}-1)^2}}
q^{9/4} \left({{t}\over{t_c}} - {{(\eta_{c0}-1)}\over{\eta_{c0} \nu}}\right)
\right)^{4/5} r_c(0) \label{r2simtext}
\end{equation}
\begin{equation}
{{dm_c}\over{dt}}=\left({{m_{c0}}\over{t_c}}\right)
\left({{r_c(t)}\over{r_{c0}}}\right)
{{6\nu}\over{ \eta_{c0}}}  \label{mdot2text}
\end{equation}
\begin{equation}
m_c(t>t_s) = m_c(0) \left({{r_c(t)}\over{q r_c(0)}}\right)^{9/4}
\label{m2simtext}
\end{equation}
Here $q=(\lambda/3)^{1/3}$, $r_s \approx
qr_c(0)(1-3(1+q)/(\eta_{c0}-1))^{4/9}$ and $t > t_s = r_c(0)/c_{PDR}$.
The PDR/core interface propagates inward as the warm gas evaporates and, finally,
the entire core is transformed to heated, photodissociated, expanding PDR material.
Using equations~(\ref{r2simtext}) and~(\ref{m2simtext}),
Figure 4 shows the clump radius and mass as a function of time for a case
where the initial column through the clump is given by $\eta_{c0} = 10$, and $\nu=5$
represents the incident FUV flux. This is a typical column density for clumps in
molecular clouds, (see \S 2) and for a clump of mass $m_c$, corresponds to
a density $n_{c0} = 3 \times 10^6 (M_{\odot}/m_c)^{1/2}$ cm$^{-3}$. In this example of
$\nu=5$, the FUV field might heat cold clump gas with initial temperature of
$\sim $30 K to about $\sim $750 K, or $\sim 10 $K gas to about 250 K.

\clearpage
\begin{figure}[f]
\includegraphics[scale=0.8,bb= 30 430 570 700]{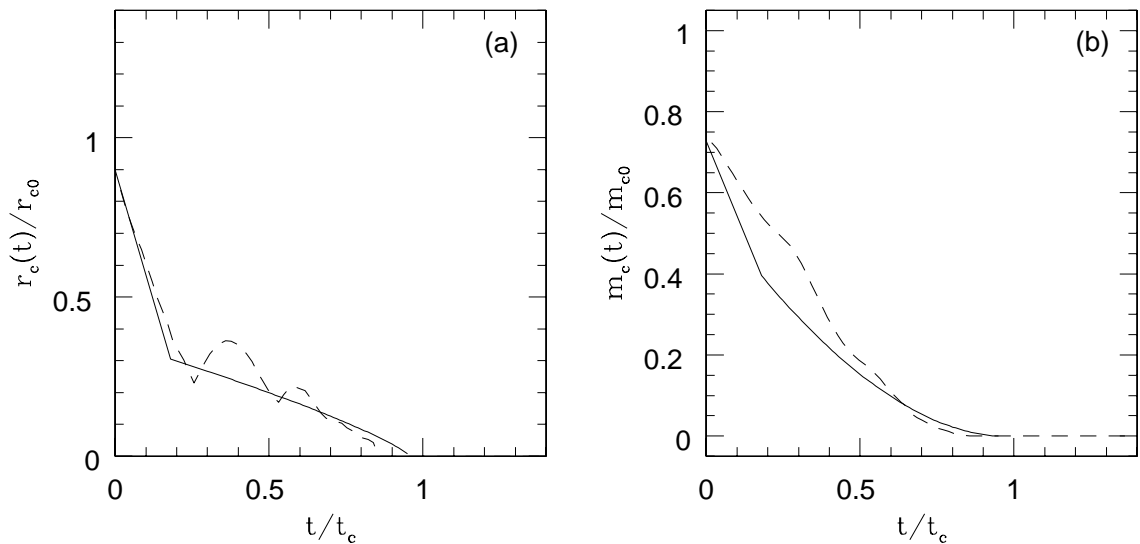}
\caption{Time evolution of radius (panel a) and mass (panel b)
of cold clump gas for a clump with $ \eta_{c0}= 10, \nu =5$ that
undergoes shock compression.  The mass and radius are
given as ratios to the initial mass and radius, and time is in units
of sound crossing times in the initial cold clump. The solid line is the analytical
result, showing an initial rapid decrease in clump radius followed by a slower
phase of clump radius evolution. The dashed line is the result from the 1-D hydrocode
and is seen to agree well with the analytical curve.}
\end{figure}
\clearpage

Also shown in Figure 4 are the results of a more detailed hydrodynamics code
simulation (described in the next section) for the same parameters. There is
very good agreement between the two results, in spite of the various simplifying
assumptions made in arriving at the analytical solutions. The numerical solution
for $r_c(t)$  oscillates with time  due to the dynamics of
the shock compression which we have neglected in the above analysis. Our analytic
solution assumes a quasi-steady state, whereas the numerical code shows the
effect of overshooting the steady state solution and rebounding.

The clump evolution goes through two distinct phases, which is apparent
from the plot of clump radius with time (Fig.[4a]). The clump initially shrinks
rapidly with time, as the shock compresses the clump in a time $t_s$.
The average velocity $v_b$ with which the radius decreases is $0.65 c_{PDR}$
for the chosen parameters. After the entire clump is compressed by the shock, it
decreases in size more slowly, now due entirely to the mass loss from its surface.
As discussed earlier, the column density in the compressed clump remains constant,
and the clump gets denser as its size decreases. In Figure 4b, the mass of the
clump is also seen to decrease with time as the entire clump gets photoevaporated
in about a sound crossing time; again, the analytical results closely follow 
the result of the numerical hydrodynamical code. The cold clump mass at $t=0$ is
less than the initial clump mass because it excludes the mass contained in the 
outer column $N_0$ which is instantaneously heated by the FUV flux
to the PDR temperature.

Using the results of the above analysis, we can obtain simple estimates of
photoevaporation timescales for a clump  exposed to incident FUV radiation.
We define the photoevaporation timescale, $t_{PE}$, as the time for the
radius of the clump to shrink to zero. The time $t_{PE}$, can be easily
determined from setting the LHS of  equation~(\ref{r2simtext}) to zero,
\begin{equation}
t_{PE} \approx 0.5 \eta_{c0}^{2/3} \nu^{-1/3} t_c \label{tpe2simtext}
\end{equation}
or
\begin{equation}
t_{PE} \approx 10^4 \left({{n_{c0}}\over{10^5  {\rm cm}^{-3}}}\right)^{2/3}
\left({{r_{c0}}\over{0.01  {\rm pc}}}\right)^{5/3}
\left({{0.3  {\rm km s}^{-1}}\over{c_c}}\right)^{2/3}
\left({{3 {\rm km s}^{-1}}\over{c_{PDR}}}\right)^{1/3}
{\rm years}\label{ay}
\end{equation}
Larger clumps are
longer-lived compared to smaller clumps in a given PDR environment.
Photoevaporation timescales of turbulent
clumps in PDRs in typical star-forming regions are thus of the order of a
few clump sound crossing times (see \S 7). 
 Paradoxically, the sudden turn-on of FUV radiation on a
clump in a vacuum can increase its lifetime, even though the FUV heats
the surface and increases the flow speed from the surface!
For turbulent clumps with lifetimes $\sim t_c$ in the absence of an FUV field,
exposure to the FUV field results in a ``pressure-confined''
period where the clump is compressed due to pressure of the heated surface PDR layer.
This compression reduces the area of the photoevaporating surface
and can extend the lifetimes of these clumps to a few $t_c$.  

\paragraph{Collapse of clumps driven by shock compression}
The FUV driven shock wave that compresses a clump to very high densities, may render it 
gravitationally unstable to collapse. Thus star formation may be triggered 
in a previously stable clump. A simple estimate of parameters leading to 
clump collapse can be made without explicitly including self-gravity in the 
equations. We use our solutions for the radius of the shock-compressed core 
(Appendix B) and compare this with its Jeans length, $r_J$. If $r_s <
r_J$, the clump is stable, otherwise it undergoes gravitational collapse.
It should be noted here that a clump supported by magnetic pressure with 
$\gamma > 4/3$ will always be stable to collapse regardless of the external 
pressure, as the magnetic pressure increases faster during compression than 
the gravitational energy density. Our discussion here is thus applicable for 
clumps with insignificant magnetic fields, or with a magnetic equation of state
where $\gamma \la 4/3$. We now solve for the collapse criterion as a function
of initial clump parameters and the FUV field strength as measured by $\nu$, for
an illustrative case where $\alpha=1, \beta=1$, and $\gamma = 1$. The radius of the
shock-compressed core is given by
\begin{equation}
r_s \simeq r_c(0) \left({{3(\eta_{c0}-1)}\over{4(\nu^2+1)}}\right)^{1/2}
\end{equation}
The Jeans length of the compressed core is given by
\begin{equation}
r_J = \left({{3 \pi c_c^2}\over{4 G m_H n_s}}\right)^{1/2}
\end{equation}
For collapse $r_s > r_J$ which yields
\begin{equation}
\eta_{c0} > 1 + \frac{3}{4}\left({{\pi n_{c0} c_c^2}\over{G m_H N_0^2}}\right)^{2/3}
(\nu^2+1)^{-1/3} \label{collapse}
\end{equation}
The factor $n_{c0} c_c^2$ in equation~(\ref{collapse}) 
is the initial thermal pressure in the clump, and for
clumps in PDRs is typically of the order of $10^{6-7}$ cm$^{-3}$ K. For
$n_{c0} c_c ^2 = 10^6$ cm$^{-3}$ K, equation~(\ref{collapse}) becomes
\begin{equation}
\eta_{c0} > 1 + {{49}\over{(\nu^2+1)^{1/3}}} \label{collapse1e6}
\end{equation}
Figure 5 shows the collapse criterion(Eq.[\ref{collapse1e6}]) on the 
$\eta-\nu$ parameter plot. Clumps with initial column densities greater
than $\sim 40 N_0$ and with initial thermal pressures of $10^6$ cm$^{-3}$ K
cannot support themselves against gravity and are unstable to collapse even 
in the absence of an FUV field. This region of instability is to the right of
the vertical dotted line in the plot. Clumps to the left of this line are 
initially stable, but sufficiently strong FUV fields may drive shocks that 
compress them and trigger collapse. The FUV fields near OB stars may thus 
trigger star formation in previously stable clumps, and {\em increase} the star 
formation rate in clouds, in cases where $\gamma < 4/3$.

\clearpage
\begin{figure}[f]
\includegraphics[scale=0.4,bb=20 145 580 710]{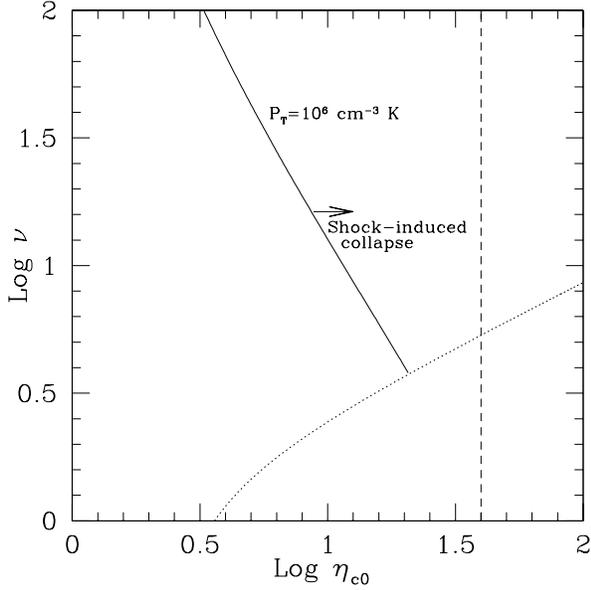}
\caption{This figure shows the $\eta-\nu$ parameter space and demarcates the 
regions where gravitational collapse can be triggered by shock 
compression (for $\gamma= 1$). The dashed line indicates the column densities 
of clumps, with thermal pressure $10^6$ cm$^{-3}$ K,
which are initially unstable to collapse. Clumps with this pressure and
columns to the left of the line are initially stable. 
The regions to the right of the solid line are where clumps
are driven to collapse by shocks induced by photoevaporation, for initial
thermal pressures in the clump $P_T= 10^6$. For higher pressures $P_{c0}$,
both the dotted ($\eta\propto P_{c0}^{1/2}$)  and solid lines 
($\eta\propto P_{c0}^{2/3}$) move to the right. }
\end{figure}
\clearpage

\paragraph{Clumps with an initial expansion phase} Clumps with initial
column densities $\eta_{c0} > \eta_{crit}$ as given by equation~(\ref{aetext})
develop very thin ($\delta_0 \ll r_{c0}$)  PDR shells when they
are first exposed to FUV radiation. This very thin shell is initially at a
high pressure and drives a shock into the clump on the inside edge of the shell
while the outside edge of the shell expands at roughly $c_{PDR}$
just as is the case for shock-compressed clumps. However, the
shock stalls before it reaches the clump centre as the PDR pressure rapidly
declines due to the shell expansion. The PDR pressure and clump pressure
become equal before the shell thickness is comparable to the clump radius.
We summarize below the detailed evolution of initially expanding clumps;
details appear in Appendix C.

Because of the high PDR sound speed and relatively small thickness, 
the PDR pressure initially drops faster than the pressure in the cold clump. The
pressure in the cold clump eventually becomes higher than that in the surrounding expanding
PDR shell, and the cold clump gas expands at its sound speed, $c_c$ (as
it would in the absence of FUV radiation since the interclump medium is
assumed here to be a vacuum). This  expansion takes place until a time $t_e$ 
at which the density (and pressure) in the clump gas
drops to that of the outer PDR layer.
For $t > t_e$, FUV photons begin penetrating into the
formerly shielded cold clump gas.  The warm PDR gas now confines the clump gas,
and further expansion of the cold clump is halted. The cold clump  proceeds to
lose mass gradually and shrinks due to photoevaporation, with an evolution similar
to the final solution for shock-compressed clumps. The radius and mass of the clump at 
times $t > t_e$ and for $\alpha=1,\beta \le 1, \gamma \le 4/3 $ are given by 
\begin{equation}
r_c(t>t_e) = r_c(t_e) - {{3 c_{PDR}}\over{2\nu^2+1}}(t-t_e)
\label{autext}
\end{equation}
where $r_c(t_e)=r_c(0)+c_c t_e$,
\begin{equation}
t_e = t_c \left(1-{{1}\over{\eta_{c0}}}\right) \left(
\left( {{3\nu + \eta_{c0} -1}\over{3\nu + 2\nu^2+1}}\right)^{1/2}
-1\right) \label{tetext}
\end{equation}
and
\begin{equation}
m_c(t>t_e) = m_{c0} {{(2\nu^2+ 1)}\over{\eta_{c0}}}
\left({{r_c(t)}\over {r_{c0}}}\right)^2 \label{avtext}
\end{equation}
Figure 6 shows the change in radius and mass of a clump with initial parameters
$ \eta_{c0} = 100, \nu =3$. The clump expands out initially and loses a small
fraction of its mass during this phase. After its density has dropped, so that
there is pressure equilibrium between the cold clump gas and the heated layer
immediately surrounding it, expansion is halted. As photoevaporation causes
mass loss off its surface, the clump shrinks in size, and is completely heated
and photodissociated in about 4 crossing timescales. The numerical results are 
also overlaid on the plot, and there is reasonable agreement.
The photoevaporation timescale for initially expanding clumps  can be defined
analogously to shock-compressed clumps as (see Appendix C, Eq.(\ref{tpeexp}))
\begin{equation}
t_{PE} \approx   \frac{4}{3}  \nu t_c  \approx
10^5  \left({{c_{PDR}}\over{3 {\rm km s}^{-1}}}\right)
\left({{0.3  {\rm km s}^{-1}}\over{c_c}}\right)^{2}
\left({{r_{c0}}\over{0.01 {\rm pc}}}\right) {\rm years} \label{az}
\end{equation}

\clearpage
\begin{figure}[f]
\includegraphics[scale=0.8,bb= 30 430 570 700]{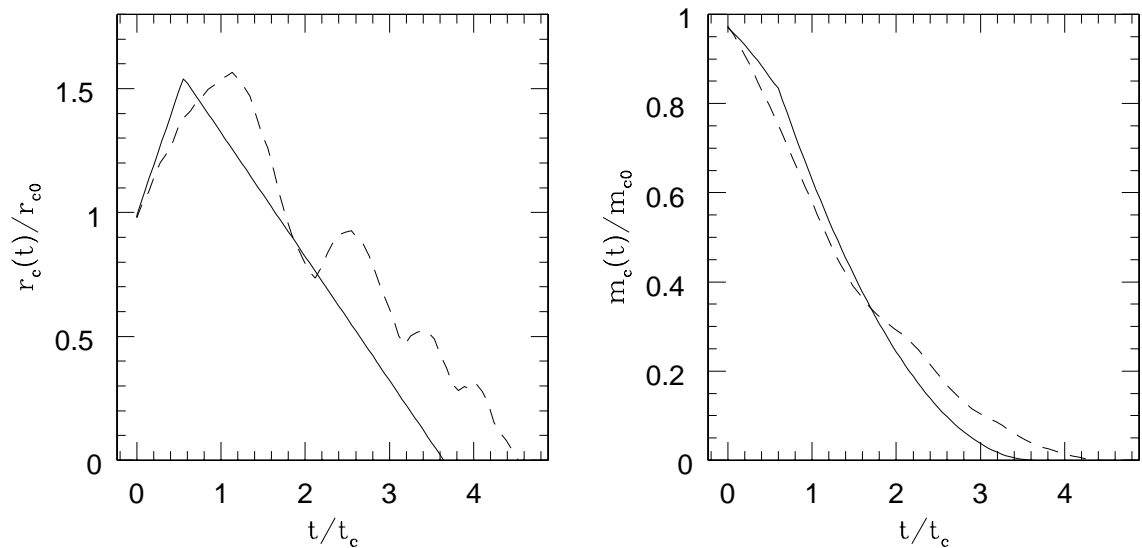}
\caption{Time evolution of radius and mass for an initially expanding clump,
with $ \eta_{c0} = 100, \nu =3$. Radius and mass are given as ratios of their
initial values and time is in sound crossing time units, $t_c$. The figure shows
the analytical result (solid line) and that from the numerical code (dashed line).
The clump radius increases in the beginning as the clump expands and then shrinks
as the clump loses mass due to photoevaporation. }
\end{figure}
\clearpage

Therefore in this regime of low $\nu$, the photoevaporation timescale
$t_{PE}$ is of order several sound crossing timescales, and is proportional
to $ \nu$.

\subsection{Analytical model for clumps pressure-confined by an interclump medium}
For a cloud model in which clumps are structures in pressure equilibrium with an
ICM, the evolution of FUV-heated clumps is somewhat different. The heated
shell or PDR can now no longer expand indefinitely, but is eventually
confined by the pressure of the ICM. The evolution of the clump is again
determined by $\eta_{c0}$ and  $\nu$.
Clumps with an initial column density $\eta_{c0} < N_0$, are completely heated 
and photodissociated instantly. Such clumps expand until the PDR gas
reaches the interclump pressure. Clumps with higher initial column densities
again are only heated on their surfaces to a column $N_0$, and their further
evolution depends on the turn-on time of the FUV field $t_{FUV}$, relative to $t_c$.

If the clumps are heated impulsively, $t_{FUV} \ll t_c$, pressure-confined
clumps initially evolve similarly to the turbulent clumps discussed earlier. 
Clumps with $\eta_{c0} < \eta_{crit}$ are thus shock-compressed, and mass
flows are set up which photoevaporate the clump.  However,
the outflow runs up against the interclump pressure and eventually reaches
pressure equilibrium with its surroundings. During this time, the clump
may either be completely heated through and get transformed into a 
sphere of warm PDR gas, or may only be partially photoevaporated. 
The final configuration for partially photoevaporated clumps consists of
a small remnant cold clump, and an extended warm PDR envelope surrounding it
and protecting it from the FUV flux. The clump, PDR envelope and interclump
medium are in pressure equilibrium. Clumps with  $\eta_{c0} > \eta_{crit}$ 
are not much affected by the FUV field. There is no expansion of the clump
due to the presence of the confining ICM, and the initially heated thin PDR shell
expands to form an expanded, but still thin, protective PDR layer.
 Containment of the PDR shell by the ICM 
prevents further penetration of FUV photons into the cold clump and there is 
no additional heating. For these clumps, photoevaporation is relatively 
unimportant and they retain large fractions of their initial masses.

Clumps that are heated with $t_{FUV} > t_c$ evolve quasi-statically, steadily
adjusting themselves to the pressure of the heated PDR shell. The evolution of
these clumps can be determined from simple steady-state equilibrium considerations. 
In their final configuration, the clumps may retain a fraction of their 
initial cold gas, surrounded by a warm PDR envelope, in pressure equilibrium with
the ICM. We consider two extremes in density profiles of the clump, a constant
density clump and a truncated isothermal sphere ($n \propto 1/r^2$)
 in pressure equilibrium with 
the ICM. 

First, we solve for the evolution of a clump of constant density. 
From conservation of mass,
\begin{equation}
n_{c0} {r_{c0}}^3 = n_f {r_f}^3 + n_{PDR}(r_{PDR}^3 - r_f^3)\label{ba}
\end{equation}
where $n_f$ and $n_{PDR}$ are the number densities in the final cold clump 
core and the PDR, and $r_{c0}, r_f$, and $r_{PDR}$ are the radii of the initial
clump, remnant clump, and the PDR envelope respectively. As the final
configuration is a  remnant clump  in pressure equilibrium with the
PDR and the ICM, the final density of the core $n_f = n_{c0}$, and
\begin{equation}
n_{c0} {c_c}^2 + \alpha n_{c0} {c_c}^2 + \beta n_{c0} {c_c}^2    = n_{PDR}
{c_{PDR}}^2  +
\alpha n_{PDR} {c_c}^2 +
\beta n_{c0} {c_c}^2 (n_{PDR}/n_{c0})^{\gamma} \ (= {\rm{ Pressure\ in\ the\
ICM}})\label{bb}
\end{equation}
Further, we know that the column density through the PDR envelope is $N_0$,
as the remnant clump is shielded from heating. Therefore,
\begin{equation}
n_{PDR} (r_{PDR} - r_f) = N_0\label{bc}
\end{equation}
Using equations (\ref{ba}) and (\ref{bc}) to eliminate $n_{PDR}$ and $r_{PDR}$,
equation~(\ref{bb}) can be written as
\begin{equation}
\nu^2 = (1+\alpha +\beta)\xi - \beta \xi^{1-\gamma} - \alpha \label{etanupc}
\end{equation}
where
\begin{equation}
\xi = \left(\eta_{c0}^3 - \eta_f^3 -\frac{3}{4}\eta_f^2\right)^{1/2} - 
\frac{3}{2} \eta_f
\end{equation}
and $\eta_f = n_{c0}r_f/N_0$. For complete photodissociation $\eta_f=0$,
or $\xi= \eta_{c0}^{3/2}$. For the standard case $\alpha=1$, $\beta=1$ and
$\gamma=4/3$, and for $\eta_{c0} > 1, \nu>1$, equation~(\ref{etanupc})
can be used to define a critical column density for complete photoevaporation. 
\begin{equation}
\eta_{cr} \simeq 0.48 \nu^{4/3} \label{etacritpc}
\end{equation}
All clumps with an initial column density lower than $\eta_{cr}$ are thus
eventually completely photodissociated. Alternately, a clump with an
initial column density $\eta_{cr}$ has to be exposed to an FUV field
($\nu$) greater than or equal to that given by equation~(\ref{etacritpc}) to
be completely photodissociated. 

\clearpage
\begin{figure}[f]
\includegraphics[scale=0.4,bb= 20 145 580 710]{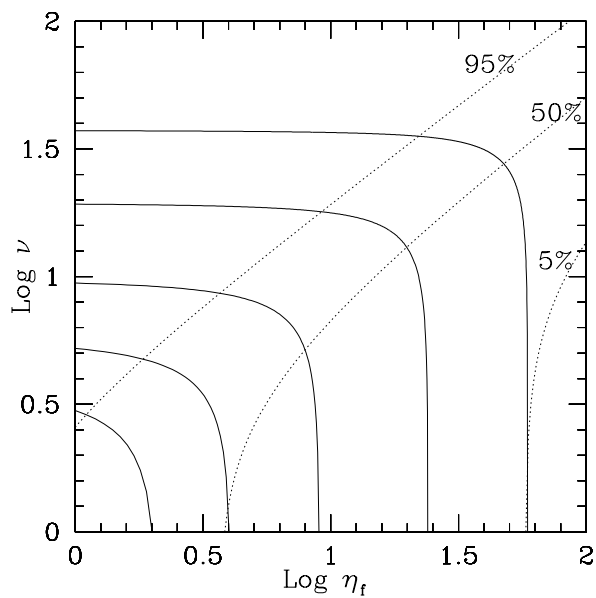}
\caption{The evolution of column density $\eta_f$ of the clump against
the parameter $\nu$.  As soon as the clump is exposed to an FUV field, 
an outer column $N_0$ is instantaneously heated, and thus for $\nu$ 
marginally greater than $1$, 
at $t=0$, the final column density $\eta_f=\eta_{c0}-1$. The figure shows
how the final column density of the equilibrium clump configuration
decreases with an increasing FUV field. Also shown are the contours 
depicting the percentage of mass lost from the clump as it photoevaporates.}
\end{figure}
\clearpage

Figure 7 shows a plot of the final column density of a clump, $\eta_f$,  
as it is exposed to various FUV fields, measured by $\nu$, for $\alpha=1$, 
$\beta=1$ and $\gamma=4/3$. The figure 
shows trajectories followed by the clump column density as the FUV
field is increased, as might happen when the clump slowly emerges
into the PDR region from the shielded interior of the molecular cloud. 
Note that the abscissa is $\eta_f$ and not the initial column density
$ \eta_{c0}$. The outer column of cold clump gas $N_0$ that is initially
heated is thereby discounted, and as soon as the FUV field is
``turned-on'' at time $t=0$, $\eta_f = \eta_{c0}-1$. 
As the FUV field ($\nu$) increases, the clump begins to lose mass due to heating
of the outer shell, and its column density begins to decrease  
with increasing $\nu$. As $\nu$ approaches the critical value for complete
photodissociation for that particular $ \eta_{c0}$,
the column density rapidly decreases with $\nu$
and the clump photoevaporates completely. If the final strength of the
FUV field is less than that required for complete photoevaporation, the
clump column density remains fixed at a point on its trajectory, as 
determined by the value of the parameter $\nu$. Therefore, given a clump's initial
column density and the strength of the local FUV field, the final column 
density of the clump can be determined. 
Equivalently, the observed $\eta_f$ and $\nu$ can be used to determine the 
initial column $\eta_{c0}$ of the clump.  The total mass loss
of the clump during its evolution can  also be evaluated from 
equation~(\ref{etanupc}). Figure 7 also shows the contours for
the mass loss being 5\%, 50\% and 95\% of the initial mass, for
different values of $\eta_f$ and $\nu$.

The evolution of a clump with an isothermal density profile can be determined
analogously. The initial density distribution in the clump is given by
$n(r)=n_{cs0}r_{c0}^2/r^2$, where $n_{cs0}$ is the density at the clump surface,
and the initial mass of the clump is $4\pi m_H n_{cs0} r_{c0}^3$. We assume that
the density in the heated PDR shell in the final configuration of the clump
is constant. Utilizing equations(~\ref{ba}-\ref{bc}) for this case, we obtain
an equation very analogous to equation~(\ref{etanupc}) where now 
\begin{equation}
\xi = \left(3\eta_{cs0}^3 - 3\eta_f^3 -\frac{3}{4}\eta_f^2\right)^{1/2} - 
\frac{3}{2} \eta_f \label{xidef_iso}
\end{equation}
$\eta_{cs0}=n_{cs0}r_{c0}/N_0$ and $\eta_f = n_{cs0}r_f/N_0$. An effective average
column density through the clump can be defined by the ratio of the mass to
area of the clump, ${\bar{\eta}}_{c0}=m_{c0}/(\pi r_{c0}^2 m_H N_0)=4\eta_{cs0}$.
The evolution of a clump with an isothermal density profile is thus qualitatively
similar to that of a constant density clump, but with a different effective
column density for the same mass. These solutions apply to clumps that 
are massive enough for gravity to be important in determining their
density structure, and therefore have appreciable central density 
concentrations. 

%
\section{Numerical code and results}
In the above analysis, simplifying approximations were made, which are
now relaxed in a numerical code to obtain complete and more exact solutions
(see Appendix D for details). One of the major differences between the
numerical code and most of the analytic equations is the inclusion of gravity in
the numerical code.
\subsection{Impulsively heated clump in a vacuum}
We first discuss the results of a numerical simulation of a clump generated by
turbulence and heated impulsively. This case is hereafter referred to as Case
 V (for ``vacuum'').
A vacuum boundary condition is used so that the clump is free to expand, even in
the absence of any heating. Initially, the thermal, turbulent and magnetic
pressures in the clump are all assumed to be equal ($\alpha=\beta=1$), and the
magnetic pressure scales with the density to the power $\gamma=4/3$. The FUV
field is  turned on instantly ($t_{FUV}$=0). The clump  is assumed to be a
constant density sphere, with a column density given by $\eta_{c0}=10$, and in
a FUV field characterized by $\nu=10$ in the outer layers. The value of
$\eta_{c0}$ corresponds to that typical of clumps in clouds, for example, $m_c=0.8
M_\odot, n_c = 2 \times 10^5 {\rm cm}^{-3}, {\rm and\ } r_c = 0.03 {\rm pc}$.
Clumps exposed to FUV fields with $G_0 \sim 10^5$ are heated to $T_{PDR} \sim 1000 
$K on their surfaces; if their shielded centres are $T_c \sim 10 K$, then the 
ratio of sound speeds $\nu \sim 10$. These initial conditions imply the clump lies in
the shock compression regime of Figure 2, and we expect from our analytic solutions that it 
should photoevaporate in less than one sound crossing timescale
$t_c$ (from eq.~[\ref{tpe2simtext}])

\clearpage
\begin{figure}[f]
\includegraphics[scale=0.6,bb=20 145 580 710]{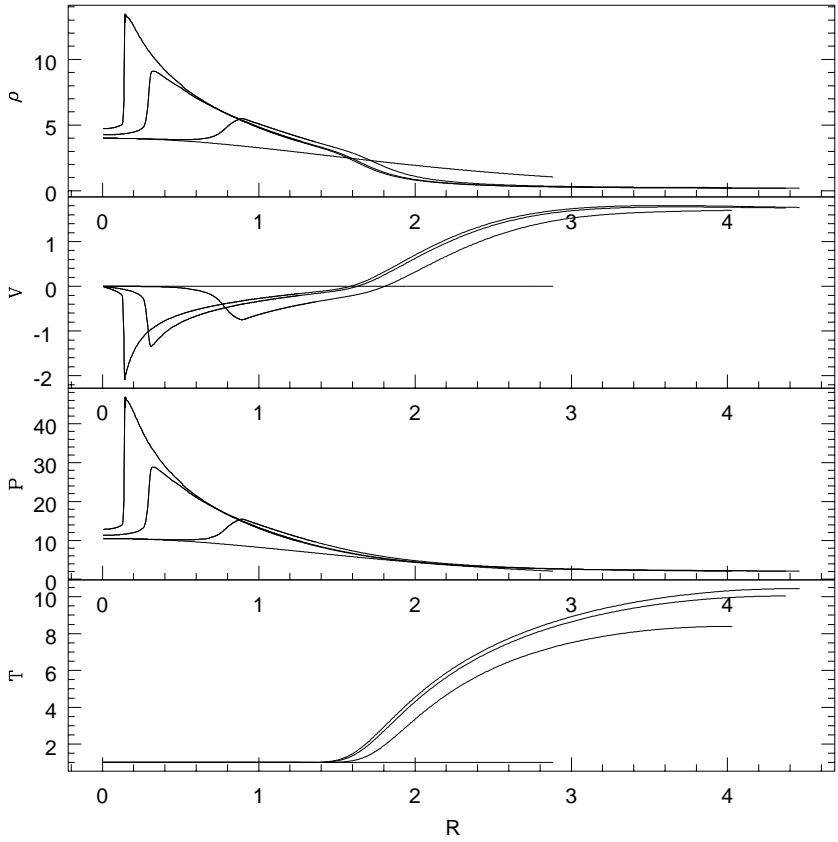}
\caption{Density, velocity, pressure and temperature as a function of radius at
times $t=0.05,0.1$, and $ 0.15 t_c$. The density, pressure and temperature are
scaled to the initial values in the clump and the velocity is in units of
the clump thermal speed $c_c$.}
\end{figure}
\clearpage

Figure 8 shows the density, velocity, pressure and temperature profiles in the
gas as a function of radius at three different instants of time, $t= 0.05t_c,
0.1t_c$ and $0.15 t_c$. As the outer layers heat up and the increased thermal pressure
causes them to expand, a shock propagates into the clump, compressing the cold
inner clump gas. The shock strengthens as it progresses to the centre of the
cloud, as evidenced by the increasing density and velocity at the position
of the shock front. The central cold gas gets compressed to very high
densities, mainly due to the convergence of the radially-moving shocked gas.
In Figure 8 we only show the evolution prior to the shock reaching the centre of
the cloud. After the shock reaches the centre, the compressed gas rebounds and the cold
inner clump undergoes radial oscillations as it settles into pressure
equilibrium with the warm, expanding PDR outflow in the outer layers.  The  mean density
of the compressed clump gas increases with time, as expected from our analytical
solutions (eq.~[\ref{al}]). As the clump loses mass, the FUV penetrates
deeper and the clump is completely photoevaporated in $0.7 t_c$. Our analytical
solution (eq.~[\ref{tpe2simtext}])
predicts a photoevaporation timescale of $0.5t_c$, which agrees
very well with the numerical solution. The photoevaporation timescale derived
for a clump with this column density, $\eta_{c0}=10$, and density, radius
and sound speed given by $n_c = 2 \times 10^5 {\rm cm}^{-3}, r_c = 0.03 {\rm
pc}, {\rm and\ } c_c = 0.3 {\rm km s}^{-1}$ respectively, is $10^4$ years.

\subsection{Gravitational collapse}
The compression of a clump by strong shocks due to FUV heating can raise the
central densities by large factors. This, however, did not render the clump
gravitationally unstable to collapse in the above case 
because $\gamma$ was chosen to be $4/3$, which raises the
magnetic pressure in dense gas sufficiently to prevent gravitational collapse
(Chandrasekhar 1961). The scaling index of the magnetic pressure with density in
clumps depends on the orientation of the magnetic field in the clump, and on
whether the field is frozen into the neutral gas. For a frozen, uniform,
unidirectional magnetic field, $P_B \propto n^{4/3}$. The magnetic field
configuration in clumps is considerably more complicated (Ward-Thompson et al.
2000), and ambipolar diffusion may allow the neutral gas to slip past magnetic
flux lines; making it difficult to define $\gamma$ unambiguously.

We investigate the possibility of triggering gravitational collapse in clumps through a
simulation where we adopt $ \gamma=1$, and all other parameters identical to
those in Case V. Such a clump is expected to undergo gravitational collapse based on our
earlier analysis (eq.~[\ref{collapse1e6}] where $\eta_{c0}=10$, and Figure 5).

\clearpage
\begin{figure}[f]
\includegraphics[scale=0.4,bb=20 145 580 710]{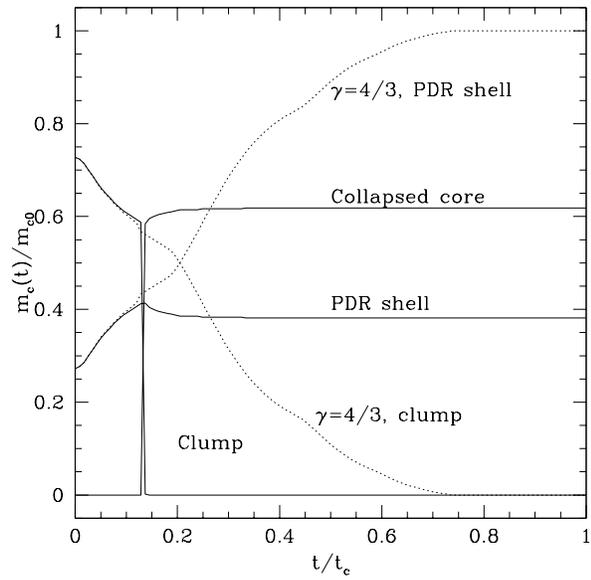}
\caption{This figure shows the mass in the clump, core and warm PDR shell as a
function of time. At collapse, the clump mass drops sharply to zero. Also shown
in dotted lines are the clump
and PDR shell masses for the case where $\gamma = 4/3$. }
\end{figure}
\clearpage

The evolution of the clump is initially similar to that in Case V. The outer
surface expands, and a shock is driven into the clump centre. As the shock
reaches the centre, the density of the cold clump gas increases and the clump
radius decreases, until the central region becomes gravitationally unstable.
A collapsing ``core'' is thus formed. This can be seen in Figure 9, which shows
the mass of the clump as a function of time. The mass of the collapsing core is
also shown in the figure, along with the results of Case V. The core mass is seen
to increase above that contained in the clump just before collapse as the core 
 accretes mass from the warm PDR gas. A significant fraction ($60\%$) of the
initial clump mass is driven gravitationally unstable to collapse, on very short
timescales. In the absence of sufficient magnetic field support, exposure to
strong FUV fields can thus trigger star formation in clumps. We also conducted a
simulation with a weaker FUV field ($\nu=5$), and found that the central regions
collapsed in that case too, but the collapsing core contained only about $10\%$
of the initial clump mass. Although these triggered collapse solutions are
interesting and instructive, we emphasize that for initially stable
clumps with $\beta \approx 1$, they hold only for $ \gamma<4/3$,
which requires ambipolar diffusion. Since ambipolar diffusion timescales are
longer than the shock compression timescale, it is likely that realistic clumps
require $ \gamma \ge 4/3$ and that triggered collapse does not often occur.

\subsection{FUV heating of a pressure-confined clump:  $t_{FUV}\gg t_c$ }
Pressure confined clumps
are modelled as Bonnor-Ebert spheres which are in hydrostatic equilibrium. We
construct a Bonnor-Ebert sphere by setting up a density distribution for a clump
with thermal, magnetic and turbulent pressure support and in hydrostatic
equilibrium. The density in the clump decreases as $r^{-2}$ in the outer regions
(similar to an isothermal sphere distribution), and flattens out in the central
regions.  The ratio of central to surface density $x$, determines the stability of
the sphere against gravitational collapse, with a value exceeding 14.4 being
``critically unstable'' for an isothermal, non-magnetic gas. 
In these runs with a confining
ICM, the outer 200 zones are used to represent the ICM. The ICM has the same
pressure as the surface of the clump and is a constant density medium with a
density (and temperature) contrast of 1000. As discussed in \S 4, for clumps
confined by an ICM, a slow turn-on of the FUV field may be more
appropriate. The FUV field in the numerical run is turned on in a time
$t_{FUV} = 5 t_c$, and we increase the temperature 
 on this timescale so that the heated outer
shell reaches its maximum temperature at $5 t_c$.   

Figure 10 shows the results for a case with varying $x$ and  where $\eta_{c0}=10, \nu=5,
\gamma= 4/3$, and $\alpha=\beta=1$. For Bonnor-Ebert spheres with $x \ga 6$, 
the outer region with an isothermal density profile contains more than half the
mass, and the clump evolution resembles that of an isothermal sphere. In the figure,
the mass of the clump is shown with time along with the analytical results for an
isothermal sphere and a constant density sphere with the same initial column. The numerical
results for three values of the central to surface density ratios, $x=1, 3$ and $6$ 
are indicated. For central concentrations greater than that represented by $x=6$, the
analytical results for an isothermal pressure-confined sphere match the
numerical results, and the agreement increases as $x$ increases and the
Bonnor-Ebert clump configuration approaches a $1/r^2$ density profile. For $x=1$, the
clump is a constant density sphere, and the corresponding analytical calculations
apply. A slight discrepancy can be noted in the figure between the two
results, and this is due to the
differences in the more realistic numerical model, i.e. an exponential drop in 
the PDR shell temperature with column and the inclusion of gravity. For intermediate
central density concentrations, as represented by the case with $x=3$, the mass loss
rate is in between the two analytical extremes, as expected. 

\clearpage
\begin{figure}[f]
\includegraphics[scale=0.4,bb=20 145 580 710]{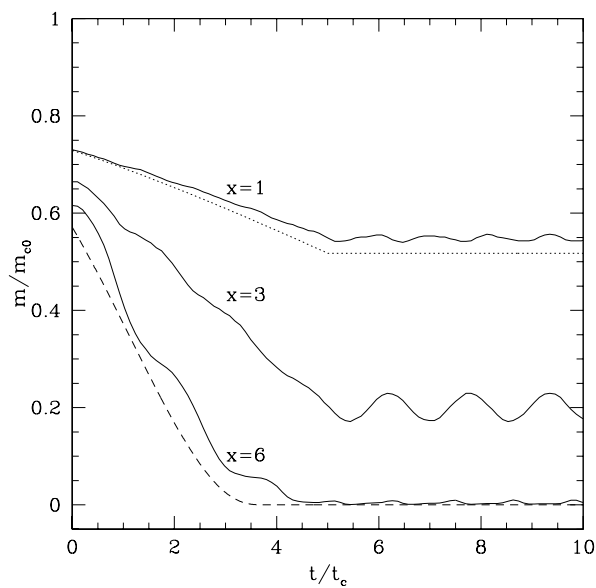}
\caption{This figure shows the evolution of mass with time  for
a slowly heated, pressure-confined clump (Bonnor-Ebert sphere)
 with varying central
to surface density ratios, $x=1,3$ and $6$. In this case,
$\eta_{c0}=10, \nu=5$, and $t_{FUV}=5 t_c$. The corresponding analytical
solutions for a constant density sphere (dotted line) and an isothermal
sphere (dashed line) are also shown for comparison. }
\end{figure}
\clearpage

%
\section{Acceleration of clumps due to rocket effect} Spherically 
asymmetric mass flows from photoevaporating clumps can cause a rocket effect
on the clump, as first noted by Oort \& Spitzer (1955) in their
theory on cloud acceleration by ionizing radiation from OB stars.
Clumps exposed to FUV radiation in PDRs see a stronger incident flux
of photons from the direction of the OB stars, though a significant
flux due to backscattered radiation from dust on the opposite surface
tends to reduce this asymmetry (see \S 5.2).  The front and back
hemispheres of the clump (with respect  to the OB star) are thus heated
to different temperatures (different  $\nu$) and
the escaping material has a higher flow velocity on the star-facing
surface of the clump. The net thrust on the clump accelerates it
away from the star, and is known as 
the rocket effect. With the help of our
analytical expressions for the evolution of an impulsively-heated,  photoevaporating clump
in a PDR, we now evaluate a simple criterion which determines the clump
size at which the rocket effect becomes significant.

A simple 1-dimensional formulation for the acceleration due to the rocket
effect is given by
\begin{equation}
{{dv_R}\over{dt}} = -{{1}\over{m_c}}{{dm_c}\over{dt}} (v_{ff} - v_{fb})
\label{momgain}
\end{equation}
where $v_{ff}$ and $v_{fb}$ are the flow velocities at the front and back
surfaces respectively and are assumed equal to the PDR
sound speed $c_{PDR}$ at those surfaces. We first compute the acceleration
due to the flow from each hemisphere separately, take the difference to
estimate the net acceleration, and then calculate the net
velocity $v_R$ attained by the clump. The mass loss from a hemisphere is
given by $2 \pi {r_c(t)}^2 \rho_b c_{PDR}$, where $\rho_b$ is the density
at the base of the flow. Taking only the component of the momentum in the
flow along the direction to the star, we obtain
\begin{equation}
\left({{dv_R}\over{dt}}\right)_{1/2}
= {{c_{PDR}}\over{ m_c}}  \pi {r_c(t)}^2 \rho_b c_{PDR}
\end{equation}
The net acceleration on the clump is therefore given by
\begin{equation}
{{dv_R}\over{dt}} = {{\pi r_{fc}^2}\over{m_c}}\left[(\rho_b c_{PDR}^2)_f
-(\rho_b c_{PDR}^2)_b\right]
\end{equation}
where the suffixes $f$ and $b$ denote the quantities for the front and back 
surfaces of the clump, and the clump radius $r_{fc}$ is determined
by the dominant flux on the front surface. 
As the initial expansion/compression phases in the evolution of a
photoevaporating clump are typically short (see \S 5), we ignore this
initial phase of evolution and assume the clump is evolving such that
its column density is at its critical value $\eta_{crit} \sim 4 \nu^2/3$,
for the standard values of $\alpha, \beta$ and $\gamma$ (see 
eq.~[\ref{aetext}]).
Along with the relation for the column through the flow $n_b(t) r_c(t) = 2 
N_0$ for each surface and the mass of the clump $m_c = 4 \pi m_H n_{fc} r_{fc}^3/3$,
(see appendix, eq.~[\ref{aa}]), we thus obtain for the net acceleration of the
clump, 
\begin{equation}
{{dv_R}\over{dt}} = \frac{9}{8} {{c_c^2} \over{r_c}}\left[1-{{(c_{PDR}^2)_b}
\over{(c_{PDR}^2)_f}}\right] \label{rocket}
\end{equation}
The clump will remain in the PDR of the GMC and be subject to photoevaporation if the
rocket velocity $v_R$ attained in a time $t_{PDR}=X_{PDR}/v_{IF}$ is less
than $v_{IF}$, or equivalently if
\begin{equation}
 r_c > \frac{9}{8} {{c_c^2}\over{v_{IF}^2}}\left[1-{{(c_{PDR}^2)_b}
\over{(c_{PDR}^2)_f}}\right] X_{PDR} = r_R \label{rocketsize}
\end{equation}
There is thus a critical size $r_R$ of clumps, where for $r_c < r_R$
they get rocketed back into
the shielded molecular cloud, and for $r_c > r_R$ remain in the
PDR and get photoevaporated or survive to photoevaporate in the \ion{H}{2} region.
 For example, for front and back surface flow 
velocities of 3 km s$^{-1}$ and 2 km s$^{-1}$ respectively, $c_c=0.3$ km s$^{-1}$
and $v_{IF}=0.5$ km s$^{-1}$ equation~(\ref{rocketsize}) 
gives a critical clump size $r_R = 0.225 X_{PDR}$.
Clumps much smaller than this will get rocketed back into the molecular cloud
as soon as they enter the PDR, partially evaporating in the process. We
emphasize again that the rocket effect is significant only for turbulent clumps
which undergo impulsive heating and photoevaporative flows. Pressure-confined
clumps are gradually heated on their exteriors and do not go through phases
of evolution with rapid mass outflows, and hence experience an
insignificant rocket effect.

The mass lost by a clump before it gets accelerated out of the PDR can be
estimated by integrating equation~(\ref{momgain}) with time, where only the 
components of velocity and mass flow along the direction towards the source 
are to be considered. Thus the mass of the clump is approximately given by
\begin{equation}
m_c(t) = m_{c0} \exp(-2 v_{IF}/\Delta c_{PDR})
\end{equation}
where $\Delta c_{PDR}$ is the difference in the flow velocities on the two
sides of the clump. For typical values of  $\Delta c_{PDR}= 1$ km s$^{-1}$,
and $v_{IF}= 0.5$ km s$^{-1}$,  in the time it takes for $v_R$ to reach
$v_{IF}$, the clump loses about $63$\% of its initial mass. Therefore clumps
which are rocketed back into the cold molecular cloud interior also lose
significant fractions of their mass before they attain high enough 
velocities to keep up with the advancing ionization front. 
%
\section{Discussion}
We begin our discussion by summarizing our results for both the case
where clumps are produced by turbulence
in regions with insignificant interclump pressure and the case where clumps
are initially stable structures confined by interclump pressure.

Consider first the scenario where clumps are continually generated by
turbulence in the molecular clouds and the PDR. Initially, the evolution
of the clump is determined by its column and the strength of the FUV flux.
If $\lambda > 1$, or $\eta_{c0} > 4 \nu^2/3$ for $\alpha=1$, $\beta=1$ and $\gamma
=4/3$, there is a very brief period of shock compression but the shock
quickly dissipates before making it to the clump centre and the clump radius
shrinks only slightly.  Subsequently, the clump expands until $\lambda
\approx 1$, at which point the photoevaporative flow halts the expansion and
proceeds to shrink and compress the clump gas, maintaining $\lambda=1$,
or a column $n_c r_c = 4 \nu^2 N_0/3$. If $\lambda < 1$, then the shock
propagates to the centre of the clump, compressing it significantly,
followed by an evolution at $\lambda \simeq 1$ identical to that above.

During the final $\lambda=1$ evolution of the shrinking, compressing, evaporating
clump, the rocket effect may play a significant role depending on the
physical size of the clump at this time. If $r_c \la r_R$, the clumps
will be accelerated into the molecular cloud, losing a significant fraction
of their mass in the process. If $r_R< r_c \la X_{PDR} c_c/v_{IF} $, the clumps will
tend to evaporate in the PDR region. However, because in the turbulent model
clumps are forming in the PDR, including the region near the  IF,
some of these clumps will survive to enter an advancing \ion{H}{2} 
region. If $r_c \ga X_{PDR} c_c/v_{IF} $ and an \ion{H}{2} region is advancing,
then most of these large clumps will survive to be engulfed by the \ion{H}{2} 
region.

The lifetimes of turbulent clumps are of order  several sound
crossing times of the initial size of the clump
 (Eq.[\ref{tpe2simtext}] and Eq.[\ref{az}]).
 As discussed previously, this is somewhat longer than clump
lifetimes in the absence of an FUV field. Since turbulent clumps are being continuously
formed, the steady-state abundance of clumps could actually increase
in the presence of an FUV field.

Based on these results, the following predictions can be made for the turbulent
scenario of photoevaporating clumps in PDRs.
\begin{enumerate}
\item There will be an enhanced population of clumps with
columns $n_c r_c \approx 4 \nu^2 N_0/3$. Note that this implies that in comparing
this population from one PDR to another, the PDR with the lower $\nu$ (e.g., due
to lower $G_0$) will have clumps with smaller column densities.
\item There will be a population of small clumps ($r_c\la r_R$) moving
with velocity $v  \sim v_{IF}$ into the molecular cloud at $A_v \ga 1$.
\item Compared with clumps in FUV-shielded regions, the clumps in turbulent
PDRs will be smaller, denser and potentially more numerous.
\item Similarly, the clumps entering the \ion{H}{2} regions will be smaller and
denser than the molecular cloud clumps, and will have columns peaked at $4\nu^2
N_0/3$.
\end{enumerate}

We note here that in the present analysis, turbulent clumps are geometrically
idealized as constant density spheres. In reality,  turbulently generated 
structures may also be sheet-like or filamentary in nature.  Let us consider
two simple representations of these geometries, disks  and 
cylinders. For these structures to survive immediate photodissociation,
the column density through their shortest dimension has to be greater than $N_0$.
Disks with larger column densities through their midplane establish photoevaporative
flows in an FUV radiation field. When the outer boundary of the heated
gas reaches a distance equal to the radius of the disk, the flow diverges and the 
density in the flow  (refer Eq.[\ref{c}] and Eq.[\ref{aa}])
 begins to drop as $r^{-3}$ if $v \propto r$. Therefore, the
mass loss rate from  the disk is essentially similar to that
of a spherical clump of size equal to the disk radius (Johnstone et al. 1998)
Photoevaporative flows from a cylindrical structure differ slightly
from those off a spherical clump of the same radius. The density at the base of the
flow is now $\propto v^{-1} r^{-1} \propto r^{-2}$ and therefore,
the critical column density $\eta_{crit} \approx 2\nu^3/3$. The mass loss rate 
from the cylindrical clump (following shock compression or initial expansion) no
longer depends on the radius and is a constant given by $-2 \pi m_H N_0 c_{PDR} l$
where $l$ is the length of the cylindrical filament. 
 (cf. Eq.[\ref{ao}]). In summary, for turbulent clumps which are likely to
be sheets or filaments, the mass loss from a thin sheet or disk of
dimension $r$ is best modelled as a clump of radius $r$. The mass loss rate
from a disk of radius $r$ and thickness $t$ is similar to that of a sphere of
radius $r$ and is $\propto r$, but the evaporation timescale is $m/(dm/dt) \propto
r^2 t/r \propto r t$ and hence much smaller than the sphere ($m/(dm/dt) \propto
r^2$) for $t\ll r$.  Filaments of 
radius $r$ evolve somewhat differently than spheres of radius $r$; their mass
loss rates do not decline as $r$ shrinks. However, the evaporation timescales
evolve similarly for cylinders or spheres since for spheres $m/(dm/dt) \propto
r^3/r \propto r^2$, whereas for cylinders  $m/(dm/dt) \propto r^2l/r^0l \propto r^2$.

Consider next the scenario where clumps are pressure-confined structures which
are bounded by a low-density ICM, and the FUV turn-on timescale is of order the
crossing time $X_{PDR}/v_{IF}$ for clumps through the PDR. For small clumps with $r_c 
\la X_{PDR} c_c/v_{IF}$, the FUV turn-on time $t_{FUV} > t_c$, which means that the
FUV quasi-statically heats and expands an outer PDR surface on the clump, continually
maintaining a pressure $P_{ICM}$ in the PDR region and in the cold clump region.
Again, the evolution of the clump is determined by its column and the strength
of the FUV flux. If $\eta_{c0} < 0.48 \nu^{4/3}$ (for $ \alpha=1, \beta=1$, and
$\gamma \ge 1$), the clump is completely heated and photodissociated as it enters 
the PDR from the molecular cloud. If $\eta_{c0} >  0.48 \nu^{4/3} $, the 
clump shrinks to a smaller cold clump surrounded by a PDR shell.  Since 
the evolution is quasi-static, there is insignificant rocket effect on these clumps.
In this scenario, small cold clumps, protected by PDR shells, can survive passage
through the PDR and into the \ion{H}{2} region. However, the minimum initial column
for such a clump is $\eta_{c0}\approx  0.48 \nu^{4/3}$ (for $\alpha=1, \beta=1$,
and $\gamma \ge 1$).

For large clumps, $r_c \ga X_{PDR} c_c/v_{IF}$, $t_{FUV} < t_c$, which means
that shocks compress the clump and strong photoevaporative outflows are initiated
as the clump enters the PDR. Clumps with $1 < \eta_{c0} < 4\nu^2$/3, are
compressed by shocks which propagate to the centre, but they finally
relax to an analogous situation as above, i.e., a thick PDR shell surrounding
a small mass cold clump if $\eta_{c0} >  0.48 \nu^{4/3} $ and totally photodissociated
otherwise.  Clumps with $  \eta_{c0} > 4\nu^2/3$ have a brief period
of shock compression and PDR expansion, but rapidly relax to a configuration with a thin
PDR shell and most of the initial mass still in the cold shielded core. All the
large clumps experience a rocket effect. However, as shown earlier, large clumps 
with $r_c > r_R$ do not experience enough rocket acceleration to prevent 
them from entering the  \ion{H}{2} region. Therefore, in the ICM confined 
scenario, a variety of clump sizes enter the \ion{H}{2} region. However, the small 
clumps entering the \ion{H}{2} region were initially clumps of larger size and mass 
in the molecular cloud, and they enter the  \ion{H}{2} region with large protective 
PDR shells.

One can compare the abundance of clumps of various sizes in PDRs with the
abundance of clumps in the molecular cloud, under the assumption that
ICM-confined clumps are long-lived (the formation time of clumps exceeds the crossing 
time $X_{PDR}/v_{IF}$). Small cold clumps are converted to PDRs but larger cold clumps
are converted to small cold clumps with PDR shells and this partially replaces
the small clump population. Thus, there will be a significant drop in the cold clump
population with $ \eta_{c0} < 4\nu^2/3$. However, the population  $ \eta_{c0} > 4
\nu^2/3$ is little affected. Therefore, a variety of clump sizes enter the \ion{H}{2}
region, but there is a suppression of clumps with columns  $ \eta_{c0} < 4\nu^2/3$
compared to the distribution in the shielded molecular cloud. 

Based on these results, the following predictions can be made for clumps initially
confined by interclump pressure.
\begin{enumerate}
\item Compared with small cold clumps ($ \eta_{c0} < 4\nu^2/3$) in similar pressure
regions of the molecular cloud, clumps of the same column density in PDRs
will have a smaller relative population. Correspondingly, 
 small and large clumps enter the advancing \ion{H}{2} regions, but the
relative number of small clumps is suppressed.
\item Larger clumps are relatively unaffected, but do experience an initial
transient period of shock compression and photoevaporative flow. 
\item Small clumps ($ \eta_{c0} < 4\nu^2/3$) in the FUV-illuminated region will
have small masses relative to their PDR shells.
\end{enumerate}

\paragraph{} We have shown that small clumps tend to be destroyed before they are 
overtaken by the ionization front of the advancing \ion{H}{2} region. This would 
affect the number, size distribution, and structure of pressure-confined clumps 
that enter an \ion{H}{2} region, and therefore affect the
evolution of the \ion{H}{2} region (see Bertoldi 1989 and Bertoldi \& McKee 1990
for a discussion of the effect of photoevaporating clumps on the evolution of an
\ion{H}{2} region). Clumps that survive to enter the  \ion{H}{2} region will
thus have undergone significant evolution from their initial states.
These clumps could be considerably denser if they are still vigorously 
photoevaporating as they enter the \ion{H}{2} region, where they continue to 
photoevaporate (Bertoldi 1989, Bertoldi \& McKee 1990).
We will discuss the propagation of clumps into \ion{H}{2}
regions in a separate paper, with application to the Eagle Nebula.

We conclude our discussion by using our results to critique the previous
inferences of dense ($\sim 10^7$ cm$^{-3}$) FUV heated clumps in PDRs and to
propose several observational consequences of our results.

Our photoevaporation models provide arguments against the interpretation that
dense ($n\sim 10^{6-7}$ cm$^{-3}$) PDR surfaces of clumps give rise to the
observed high excitation CO lines. The fundamental problem is the propagation of
the FUV photons to regions of such high density. If densities of $\sim 10^{6-7}$
cm$^{-3}$ are in fact heated to $T\gg 100$K, then the thermal pressures in these
regions are much larger than the pressures in the interclump gas. The only way
this is possible is if: (i) the clumps are gravitationally bound or (ii) the
warm surface regions are photoevaporating, as discussed in this paper. In either
case, the density in the clump will smoothly fall off with radius, $n(r) \propto
r^{-b}$ where $b \sim 2-3$, until the transition to the ICM.
However, FUV photons which heat the CO are only able to penetrate through a column
$N_0 \sim 2\times 10^{21}$ cm$^{-2}$, where
\begin{equation}
N_0 \approx \int_{r}^{\infty } n(r)dr  \approx {{1}\over{b-1}} n(r) r
\label{a}
\end{equation}
This equation reveals the size $r$ of the FUV heated region if a density $n(r)$
is required in this region. It implies that for the FUV to heat gas at densities
$\ga 10^6$ cm$^{-3}$, the clumps must be very small with size scales of $r\approx
N_0/10^6$ cm$^{-3}  \approx 2 \times 10^{15}$ cm at this density.
Gravitationally bound clumps have densities $n\sim 10^{6-7}$ cm$^{-3}$
at sizes $\gg 10^{15}$ cm (e.g., Shu 1977). In such clumps, the FUV would
not penetrate to such high densities, but would be absorbed in the lower density
regions further out. Such small scales as $10^{15}$ cm also
argue against unbound clumps explaining the high temperature CO emission.
We have shown in this paper that such a small clump will either be
rocketed out of the PDR or photoevaporate in
roughly  a sound crossing time, or $\sim 300$ years, a time so short as to rule
out this possibility.

The size scale of 100 AU is suggestive of disks around
young stars rather than clumps, so that one might appeal to photoevaporating
disks like the ``proplyds'' in Orion (e.g., Johnstone et al. 1998).
Here, the enormous mass reservoir of the disk greatly lengthens the
photoevaporative timescale. However,
the area filling factor of these photoevaporating disks required to match
the CO high-J line intensities  in, for example, the Orion Bar region
is of order $\sim 0.1$ (Burton et al. 1990).
Such a high area filling factor requires a volume
density of $\ga 3 \times 10^5 $ stars pc$^{-3}$, more than an order of magnitude
higher than that observed in the densest region of the stellar cluster near the
Trapezium (McCaughrean et al. 1994). Photoevaporating proplyds are thereby also
ruled out.

Therefore, it seems quite  unlikely that FUV radiation
is heating dense, $n\sim 10^{6-7}$
cm$^{-3}$, PDR surfaces of clumps or that photoevaporating proplyds can explain
the mid-J CO emission. We suspect that the CO emission arises in
$n\sim 10^5$ cm$^{-3}$ surfaces of large ($\ga 10^{16}$ cm) clumps,
and that the chemistry, heating and/or dynamics needs to be
modified in PDR models in order to better match the observations.

There are several observational consequences implicit in the scenario
of a strongly photoevaporating (i.e., $t_{FUV}/t_c < 1$) clump. First, the
molecular material from the clump core is effectively advected out into
the PDR region, as has been discussed by Bertoldi \& Draine (1996),
St\"{o}rzer \& Hollenbach (1998), Hollenbach \& Tielens (1999). This can
have strong effects on the chemistry, primarily because H$_2$ can exist in
surface regions where it would have been atomic in a stationary, steady-state
case. In turn, the higher abundance of H$_2$ can chemically enhance the abundance 
of other minor species, such as H$_3^{+}$ (Bertoldi \& Draine 1996) and CO and can
modify the heating and cooling processes and therefore the PDR temperature.
Initial results (St\"{o}rzer \& Hollenbach 1998 and St\"{o}rzer 2000, private
communication) suggest that advection will enhance the intensities of the
mid-J CO lines, by creating more warm CO near the surface. Finally, the
flow will broaden line widths to values
 $\Delta v_{FWHM} \sim c_{PDR}$. However, it
should be noted that many common molecules, such as CO, HCO$^{+}$, CS, HCN,
CN still exist primarily at column depths $N > N_0$ from the surface, where the
advection velocities are low. Broader linewidths should be seen in
\ion{C}{2}(158$\micron$),\ion{O}{1}(63$\micron$) and possibly the
H$_2$ (2$\micron$) lines.

%
\section{Applications to observed PDRs in well-studied star forming regions}
We compare our results with available observational data on clump 
characteristics from some well-studied star-forming regions, the Orion Bar, 
M17SW, NGC 2023 and the PDR surrounding the Rosette Nebula. Clump column
densities ($\eta$) can be estimated from measured or inferred
densities and sizes of clumps in 
these PDRs. Densities are usually inferred indirectly from comparing
observed line intensities and line ratios with available PDR chemical models
in most of these cases. Clump sizes are  ill-determined, as they are usually 
too small to be resolved directly and only upper limits to the sizes are 
available, with the PDRs in Rosette and NGC 2023 being possible exceptions. 
The FUV field strengths in these PDRs can be estimated more accurately, as
the luminosities of the illuminating O/B stars is reasonably well-determined. 
We use these field strengths and the dust temperature models
of Hollenbach, Takahashi \& Tielens (1991) to estimate the temperature of dust
in the shielded clump interior. As the interior gas temperature is closely coupled to the
dust temperature, we set the two  equal and thus determine the sound speed
$c_c$ in the cold clump gas. The temperature to which the FUV field heats the
surface clump gas is determined from the PDR models of
Kaufman et al. (1999), and thus $c_{PDR}$ is obtained, allowing us to
estimate the value of the parameter $\nu$ in each region. Below we first
determine $\eta$ and $\nu$ for each observed region, and then use these
parameters, compared to Figures 2 and 7, to infer their nature and evolution.

\paragraph{Orion Bar} The Orion Bar has been observed in many atomic and
molecular line transitions in the sub-millimeter and infrared wavelength regions (e.g. 
van der Werf et al. 1996, Tauber et al. 1994, Young Owl et al. 2000, Marconi et al.
1998). The molecular line emission peaks at about 0.03 pc from the IF, and the
gas between this dense ridge and the IF is  mainly neutral atomic hydrogen. The
observed line intensities and line ratios  appear to be best explained by
a two-component model for the PDR gas, with small ($\la 0.02$ pc)
dense ($n_c \sim 10^{6-7}$ cm$^{-3}$) clumps,
required to match the line emission and an interclump medium $n_{ICM} \sim 10^{4-5}$
cm$^{-3}$ which causes the chemical stratification of the edge-on PDR.
The local FUV field is estimated as $G_0 \sim 4 \times 10^4$. Such an FUV
field incident on clumps with these inferred densities would heat their surfaces
to temperatures $\sim 2000-5000$ K (Kaufman et al. 1999). The gas temperature
in the interior of clumps exposed to this FUV flux is expected to be $\sim 50$ K
(Hollenbach et al. 1991). We thus obtain $\eta \sim 15-150$  and $\nu \sim 8-10$.

\paragraph{M17 SW} Multilevel molecular line observations in CS (Wang et al.
1993, Evans et al. 1987) and in several fine structure lines (\ion{C}{2},
\ion{Si}{2}, \ion{O}{1}, Meixner et al. 1992) indicate that the M17 SW
cloud core consists of numerous high density clumps with densities $n_c \sim 
1-5 \times10^5$ cm$^{-3}$. Maps of CO$^{+}$ emission compared with theoretical
models suggest that the emission is produced in the
warm surface layers of PDRs in dense clumps (St\"{o}rzer et al. 1995) of sizes
$r_{c}\sim 0.1$ pc. The local FUV field near the \ion{H}{2}
region/molecular cloud interface is $ G_0 \sim 8  \times 10^4$, which implies
PDR temperatures of about 2000K on the clump surfaces
and cold clump gas temperatures $\sim 60$K, therefore, $\nu \sim 4-9$.
The column densities of clumps are typically $\eta=7-35$, as estimated
from typical densities and sizes (Meixner et al. 1992, St\"{o}rzer et al. 
1995).

\paragraph{NGC 2023}
The well-studied PDR in the reflection nebula NGC 2023, at a distance of 475 pc, is
illuminated by a B star and is well-studied (e.g., Wyrowski et al. 2000,
Steiman-Cameron et al. 1997, Howe et al. 1991). Observations of atomic fine
structure lines indicate the presence of clumpy gas in the PDR with densities
$\sim 10^5$ cm$^{-3}$, and an ICM density of $\sim 10^3$ cm$^{-3}$ (Steiman-
Cameron et al. 1997). The PDR is situated about 0.1-0.2 pc away from the
exciting star, and has a width $X_{PDR}=0.04$ pc. The clumps have sizes 
$\sim 0.05$ pc (Wyrowski et al. 2000), and therefore $\eta_{c0} \sim 5$. 
The FUV field strength has been estimated at $G_0 = 10^4$, and from the PDR 
models of Kaufman et al. (1999), $c_{PDR}= 3$km s$^{-3}$. With a clump
gas temperature of $\sim 20$ K for this value of $G_0$, we obtain $\nu = 5$. 
It should be noted that
the level of turbulent energy in the NGC 2023 PDR as estimated from observed 
linewidths is low (Wyrowski et al. 2000), which suggests that clumps are not 
rapidly re-formed by turbulence in the PDR.

\paragraph{Rosette PDR} The Rosette Nebula surrounding the young
open cluster NGC 2244, is an expanding \ion{H}{2} region at a distance of 1.6
kpc. The \ion{H}{2} region/molecular cloud interface shows a distinct ridge of
emission, with substructure on small scales down to 0.5 pc (Schneider et al. 1998).
The PDR region is very extended, and the FUV photons penetrate deep into the
cloud with an interclump medium density $n_{ICM} \sim 10$ cm$^{-3}$ (Blitz
1991, Williams et al. 1995). The ICM near the edge of the HII region may
even by partially ionized gas as suggested by the pervasive H$\alpha$
emission observed (Block et al. 1992). This region has a low FUV field, with $G_0 \sim
200$ about 15 pc away from the OB star cluster, and dropping to lower values of
10-50, about 30 pc away from the cluster (Schneider et al. 1998).
Clump densities required to match observed CII line intensities are $ \sim
10^{4-5}$ cm$^{-3}$. Cold clump gas temperatures are calculated to
range from $10-15$ K. 
The Rosette PDR thus has $\eta_{c0} \sim 5-35$ and $\nu \sim 5$, while
further away from the \ion{H}{2} region, $\nu \sim 2-3$.

\paragraph{} Using the above determined values of the parameters $\eta$ and $\nu$ for
clumps in the above PDRs, we can locate them on the parameter plots of
Figures~2 and 7. Figure 11 shows an overlay of our model results with 
inferred data for both the turbulent and pressure-confined clump models.
The data are consistent with both  clump models and,
within the limits of the uncertainties in arriving at ``observed'' 
 $\eta$ and $\nu$, agree reasonably well with our predictions.
Turbulent clumps are expected to have columns fairly close to the
critical value $\eta_{crit}$ which is $\approx 4 \nu^2/3$. The observed data
lie near the $\eta=\eta_{crit}$ line, indicating that the observed clumps
may be turbulent in origin and undergoing photoevaporation in the FUV
field. The exception is NGC 2023. The clumps in this PDR are observed to have negligible
turbulence, and are thus not expected to have a column equal to  $\eta_{crit}$.
The data are also consistent with clumps being pressure-confined. In such a
situation, we would not expect to see clumps beyond the 95\% mass loss line,
as seen in the figure. However, there does appear to be a weak trend of
increasing $\eta$ with $\nu$, which if real is
not easily explained in this model. The clumps
in NGC 2023 are probably pressure-confined, and seem to have lost almost
75\% of their original mass through FUV heating. 
High resolution observations of clumps in PDRs, with more accurate
determinations of the sizes and densities of clumps would help in 
locating them more precisely on the $\eta-\nu$ plots, and thus 
be able to distinguish between the pressure-confined and turbulent clump
models. 

\clearpage
\begin{figure}[f]
\includegraphics[scale=0.8,bb= 30 430 570 700]{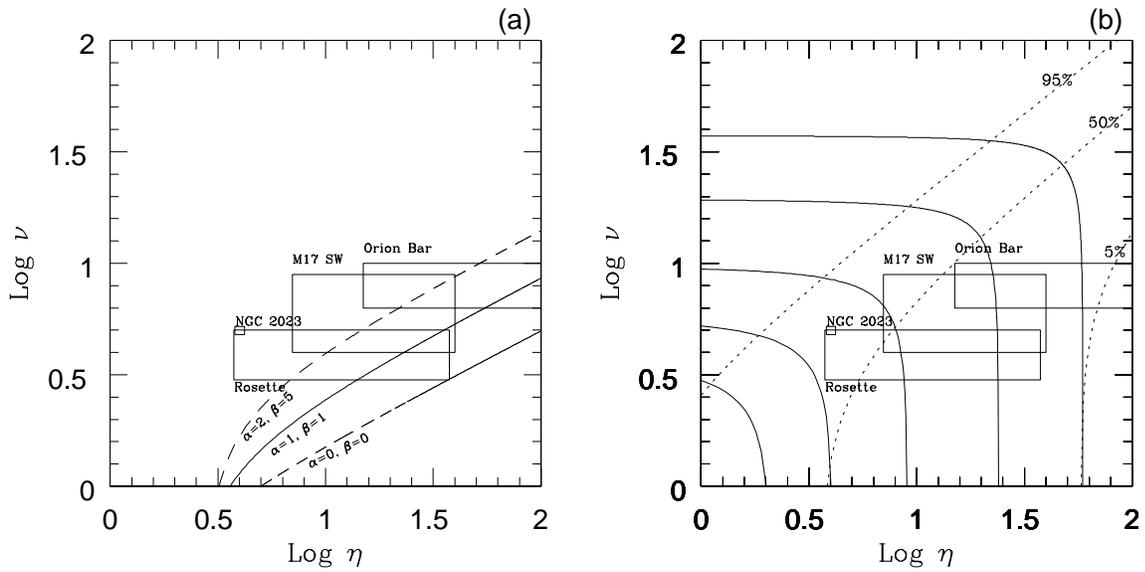}
\caption{ Observational data for some PDRs on the $\eta-\nu$ parameter plot
for the turbulent (panel a) and pressure-confined (panel b) clump models.
There is a slight indication that the columns lie along the $\nu^{4/3}$
lines expected in the turbulent model of clumps, with the exception of
NGC 2023.}
\end{figure}
\clearpage

%
\section{Conclusion}

We have studied the effects of FUV radiation from young OB stars on the
evolution of dense structures or clumps in photodissociation regions. We
find that the ambient FUV field penetrates through the surface of a
dense clump and heats this surface layer to high temperatures,
causing mass loss and thereby inducing photoevaporation of the clump.
Through analytic approximations and numerical hydrodynamical calculations,
we determined the evolution and lifetimes of clumps, subject to various
physical conditions and initial parameters.

The evolution of a clump is mainly determined by three parameters, the ratio of the
initial column density to the column penetrated by the FUV flux,
$\eta_{c0}$;  the strength of the FUV field, 
which we denote by the ratio of the sound speeds in the heated gas and the cold 
clump material, $\nu$; and the timescale for the turn-on of the FUV field, relative
to the sound crossing time  through the clump, $t_{FUV}/t_c$. We consider
the evolution of turbulent, impulsively-heated clumps and also 
pressure-confined clumps in a PDR. Impulsively heated clumps with a
photoevaporation parameter $\lambda \approx 3\eta_{c0}/(4\nu^2) < 1$
are initially compressed by a shock driven by the external high-pressure FUV-heated
layer to high densities, and later lose mass as the FUV gradually penetrates
through the entire clump. Clumps with $\lambda > 1$ go through an initial expansion
phase after the rapid decay of the shock wave. Both clumps with  $\lambda < 1$ and
$\lambda > 1$ evolve toward the $\lambda = 1$ condition. The final evolution
toward complete photodissociation and heating occurs at constant column through
the clump, the density increasing as the clump shrinks. Photoevaporation timescales
in all these cases are typically a few $t_c$, suggesting clump destruction
times of $ \sim 10^{4-5}$ years, under typical PDR conditions. Slowly
heated clumps evolve quasi-statically, and if confined by an external ICM pressure,
a fraction of the initial cold clump mass may be retained if $\eta_{c0} >
0.48\nu^2$.  Clumps with
initial column densities $\eta_{c0} < 0.48 \nu^2$ are completely photoevaporated
by the FUV flux.  

We predict that in the turbulent scenario, observed clumps should all have
columns close to the critical value for the local FUV field, $\eta_{crit} 
\approx 4 \nu^2/3$. Clumps lifetimes are prolonged by photoevaporation and
we also expect a higher steady-state abundance of clumps in PDRs as compared
to the shielded molecular cloud interior. For pressure-confined clumps, there
will be a decrease in the number of smaller clumps in PDRs, and many clumps
will lose substantial fractions of their mass to extended warm PDR shells
around them. 
We compared our results to observations of some well-studied PDRs, such as
the Orion Bar, M17SW, NGC 2023 and the Rosette PDR, and find that the
data are consistent with both clump models, but perhaps favour the
turbulent clump interpretation. Clumps entering the \ion{H}{2} region
around an O/B star with a D-type ionization front
are expected to have thus undergone significant evolution as they pass
through the PDR. 

\acknowledgments
We acknowledge support from the NASA Astrophysical Theory Program under RTOP
344-04-10-02, which supports the Center for Star Formation Studies.
U.~Gorti was supported by a National Research Council Associateship.
We thank K.~R.~Bell, M.~Kaufman, C.~McKee and H.~St\"{o}rzer  for useful discussions.

\appendix
\section{Analytic model: Unconfined clumps with $t_{FUV}/t_c \ll 1$}
We first define several characteristic timescales during evolution. The
initial sound crossing time through the clump is denoted by 
$t_{c}=r_{c0}/c_c$.
We can similarly define a sound crossing time through the warm PDR layer, as
$t_{p}=\delta_0/c_{PDR}$. Finally, we define the timescale for the PDR layer
to expand to a thickness comparable to $r_{c0}$, $t_{e0}=r_{c0}/c_{PDR}$.

We relate the density $n_b$ at the base of the PDR shell ($r=r_c$) to $N_0$
for epochs when the PDR shell has expanded to a thickness greater than the
clump
radius ($t > t_{e0}$). The FUV-heated shell of initial thickness $\delta_0$
expands as an isothermal flow into vacuum and the characteristics of the
flow regions are essentially similar to that of the Parker wind solution 
with
no gravity (Parker 1958). For simplicity, it is assumed here that once the
warm PDR region has expanded to $\delta \gtrsim r_c$ or $t>t_{e0}$, the gas
leaves the clump surface at $c_{PDR}$ and the velocity of the heated gas
increases
linearly with radius. Based on our hydrodynamical code results,
this is a reasonable approximation to within about two clump
radii, where most of the column density in the PDR gas lies. The clump mass 
loss
rate is determined by the penetration depth of the FUV flux, and hence 
$N_0$.
For a power law density profile in the shell ($n(r) \propto r^{-b}, b 
\gtrsim
2$),
most of this column arises from the base of the shell, or, more precisely, from
the region between the base (clump surface) $r_c$ and $\sim 2 r_c$. From the
equation of continuity and $ v \propto r$, the steady flow density profile of
the warm PDR shell is obtained as,
\begin{equation}
\label{c}
n(r)   \propto v^{-1} r^{-2} \propto r^{-3},
\end{equation}
valid for $r_c < r < 2r_c$ once the shell has expanded to at least two clump
radii ($t > t_{e0}$). Equation~(\ref{c}) can be easily 
integrated
over $r$ to obtain the column through the warm PDR at any given instant of
time. Let $r_c$ denote the instantaneous cold clump radius and $r_t$ the 
total
radius of cold clump and warm PDR gas. At $t=0$, $r_c= r_c(0)=r_{c0}- 
\delta_0$.
Using the fact that $r_t \gg r_c $ for $t > t_{e0}$, and as the FUV flux 
always
penetrates through a column $N_0$, we have
\begin{equation}
N_0 = \int_{r_c}^{r_t} n(r,t)dr  \approx \frac{1}{2} n_b(t) r_c(t)
{\rm \ for\ } t > t_{e0}
\label{aa}
\end{equation}

We next write down the equations of momentum flux or pressure. The pressure
in the shocked cold clump gas is the sum of thermal, turbulent and magnetic
pressures and given by
\begin{equation}
P_c(t) = n_c(t)c_c^2 +  \alpha  n_c(t)c_c^2 +
\beta n_{c0} c_c^2 (n_c(t)/n_{c0})^{\gamma} \label{ab}
\end{equation}
where $n_c(t)$ is the shocked clump density at time $t$, $n_{c0}$ is the
initial clump density, $ \alpha$ is the initial ratio of turbulent and thermal
pressures, and $\beta$ is the initial ratio of magnetic to thermal pressure 
in
the
clump. The  pressure at the base of the PDR flow, $P_b$, is given by
\begin{equation}
P_b(t)= n_b(t) (c_{PDR}^2+\alpha c_c^2) +  \beta n_{c0} c_c^2 \left(
{{ n_b(t)}\over {n_{c0}}} \right)^{\gamma} + n_b(t) v_{flow}^2  , \label{ac}
\end{equation}
where $v_{flow}$ is the outward flow at the base. We assume that $v_{flow}=
c_{PDR}$, and from equations~(\ref{aa}) and~(\ref{ac}),
\begin{equation}
P_b(t)= 2 {{N_0}\over {r_c(t)}} (2 c_{PDR}^2 +\alpha c_c^2) +
\beta n_{c0} c_c^2  \left( {{2 N_0}\over {n_{c0} r_c(t)}}
\right)^{\gamma}\label{ad}
\end{equation}

The boundary between the ``shock compression'' and ``initial
expansion'' regions  of Figure 2 is determined by
comparing equations~(\ref{ab}) and~(\ref{ad}). Values of the parameters
$\eta_{c0}$ and
$\nu$ on this boundary are such that the shock dies out just before making
it to the centre. The pressure in the clump at this instant is about the 
same as
the initial pressure and $r_c(t_{e0}) \simeq r_c(0)$. In a time $t=t_{e0}$,
the outer edge of the PDR shell expands out to $2 r_{c0}$, and the initial
PDR column begins to  decrease appreciably due to spherical divergence, 
allowing
further FUV flux penetration.

If $P_b(t_{e0}) > P_c(t=0)$, the pressure in the PDR flow is sufficiently
high for the shock wave to reach the centre of the clump, compressing the 
clump
and raising its central pressure so that $P_b(t_{e0})=P_c(t_{e0})$. Such
clumps lie in the ``shock compression''
 of Figure 2. If the initial parameters are such that
$P_b(t_{e0}) < P_c(t=0)$, the shock wave rapidly dies out before reaching
the centre and the clump then stops contracting and begins to expand at a 
speed
$\approx c_c$, maintaining pressure equilibrium. These two evolutionary
sequences are separated by the ($\eta_{c0}, \nu$) boundary obtained setting
$P_b(t_{e0}) = P_c(t=0)$, or from equations~(\ref{ab}) and~(\ref{ad}),
\begin{equation}
{{2(2\nu^2+\alpha)}\over{\eta_{c0} -1}} + {\beta} \left(
{{2}\over{\eta_{c0}-1}}
\right)^{\gamma}
  = 1 + \alpha + \beta. \label{ae}
\end{equation}
For values of $\eta_{c0}$ and $\nu$ typical in PDRs such as Orion 
($\eta_{c0}
\gg 1, \nu^2 \gg 1, \alpha=\beta=1,\gamma=4/3$) this relation can be
approximated
as $\eta_{c0} \approx 4\nu^2/3$.
A photoevaporation parameter $\lambda=3(\eta_{c0}-1)/2(2\nu^2+1)$ can be
defined such that the relation $\lambda=1$ demarcates the two regions.
Clumps with $\lambda < 1$ are shock compressed and those with $\lambda > 1$
initially expand. 
\section{Evolution of shock-compressed clumps}
 As the flow expands in a shock compressed clump,
the clump shrinks by  losing mass at a rate
\begin{equation}
{{dm_c} \over {dt}} = -4  \pi m_H n_b(t) r_c(t)^2 c_{PDR}\label{af}
\end{equation}
Initially, a shock is driven into the clump, but we do not attempt to analytically 
model the propagation of the shock wave in detail. We assume that the shock 
travels inward at the PDR sound speed and the entire clump gets compressed in a time
$t_s = r_c(0)/c_{PDR}$, where $r_c(0)= r_{c0} - \delta_0$. We first solve 
for the clump radius, mass and mass loss rate for $t<t_s$, during the epoch when 
the shock propagates to the centre.

We assume that the clump radius decreases with a constant speed $v_b
(\lesssim
c_{PDR})$ so that
\begin{equation}
r_c(t)= r_c(0) - v_b t.\label{ag}
\end{equation}
The mass of the cold clump decreases with time, and at $t_s$ is given by
\begin{equation}
m_c(t_s) = m_c(0) +  \int_0^{t_s} (dm_c/dt)dt\label{ah}
\end{equation}
where $m_c(0)= \frac{4}{3} \pi m_H n_{c0} {{r_c}^3}(0)$.
From equations ~(\ref{aa}),~(\ref{af}),~(\ref{ag}) and~(\ref{ah}), we have
\begin{equation}
m_c(t_s) = m_c(0) - 8  \pi m_H N_0 r_c(0)^2 ( 1 - v_b/(2 c_{PDR}))\label{ai}
\end{equation}

We solve for $v_b$ and thereby calculate $r_c(t_s)$. To do this we need to
calculate
the clump compression at $t_s$. The clump gets compressed by the shock to a 
new
density $n(t_s)$ which can be determined from the condition of pressure
equilibrium at the clump surface. The momentum flux conservation equation 
across
the ``front'' which marks the boundary between clump and FUV heated flow is 
then
\begin{equation}
(1+\alpha)n_c(t_s) c_c^2 + \beta n_{c0} c_c^2 (n_c(t_s)/n_{c0})^\gamma
= 2 n_b(t_s) c_{PDR}^2 + \alpha n_b(t_s) c_c^2 +
\beta n_{c0} c_c^2  (n_b(t_s)/n_{c0})^\gamma \label{aj}
\end{equation}
In moderate to strong FUV fields, the PDR sound speeds and hence the shock
velocities (assumed equal to $c_{PDR}$) are high, and the 
resulting compression is also high. As the magnetic pressure $P_B$, scales with a 
higher power of the density than the thermal pressure $P_T$, (we use $\gamma 
\gtrsim
4/3$)
and as the
initial magnetic pressure is comparable to the thermal pressure, the thermal
contribution to the pressure in the compressed gas can be ignored. In the
expanded flow, the density is low and here the thermal pressure and 
dynamical
pressure dominate. Equation~(\ref{aj})  can be simplified to give the 
density
of the compressed gas
\begin{equation}
{{n_c(t_s)}\over{n_{c0}}} = \left(
{{(2\nu^2+\alpha) }\over{ \beta }} {{n_b(t_s)}\over{n_{c0}}}  \right)^{1/
\gamma}.\label{ak}
\end{equation}
From equations~(\ref{aa}) and~(\ref{ak}),
\begin{equation}
{{n_c(t_s)}\over{n_{c0}}} = \left({{2 \delta_0 (2\nu^2+\alpha)}\over{\beta}}
\right)^{1/ \gamma} r_c(t_s)^{-1/ \gamma}.\label{al}
\end{equation}
The mass of the clump at time $t_s$ can also be expressed as
\begin{equation}
m_c(t_s) = {{4  \pi}\over {3}} m_H n_c(t_s) r_c(t_s)^3. \label{am}
\end{equation}
From equations~(\ref{ag}),~(\ref{al}) and~(\ref{am})
\begin{equation}
m_c(t_s) = {{4  \pi}\over {3}} m_H n_{c0}
\left({{2 \delta_0 (2\nu^2+\alpha)} \over{\beta}}   \right)^{1/ \gamma}
r_c(0)^{3- 1/ \gamma}  \left(1-{{v_b}\over {c_{PDR}}}\right)^{3-1/
\gamma}.\label{an}
\end{equation}
At time $t_s$, the radius of the shock-compressed clump is
$r_s=r_c(t_s)=r_c(0)(1-v_b/c_{PDR})$ from equation~(\ref{ag}).
The unknown parameter $v_b$ is finally determined from equating the two
expressions for the clump mass, equations~(\ref{ai})  and~(\ref{an}).
The evolution of the clump radius is thus known as a function of time for
$0<t \le t_s$, from equations~(\ref{ag}),(\ref{al}) and the solution for 
$v_b$.
These general equations are, however, fairly complex, and an approximate
solution can be obtained for the case $\alpha=1, \beta=1$, and $\gamma=4/3$
as,
\begin{equation}
r_s \approx q r_c(0) \left(1-{{3(1+q)}\over{\eta_{c0}-1}}\right)^{4/9}
\end{equation}
where $q= (\lambda/3)^{1/3}$.

We now solve for the evolution of the clump for $t > t_s$. After being
compressed by the shock, the clump continues to lose mass through
photoevaporation, with a  mass loss rate,
\begin{equation}
{{dm_c} \over {dt}} = -4  \pi m_H n_b(t) r_c(t)^2 c_{PDR} =  -8 \pi m_H N_0
r_c(t)  c_{PDR}
\label{ao}
\end{equation}
The mass of the clump at any given time ($t > t_s$) can also be written as
\begin{equation}
m_c(t) = m_c(0)
\left({{2(2\nu^2+\alpha)}\over{\beta(\eta_{c0}-1)}}\right)^{1/\gamma}
\left({{r_c(t)}\over{r_c(0)}}\right)^{3-1/\gamma}. \label{ap}
\end{equation}
From equation~(\ref{ao}) and differentiating equation~(\ref{ap})  with time, 
the
radius of the clump as a function of time is given
\begin{equation}
r_c(t)= \left(\left({{r_s)}\over{r_c(0)}} \right)^{2-1/ \gamma}
- {{(2\gamma-1) }\over{(3\gamma-1)}} {{6 \nu \eta_{c0}  }\over{(\eta_{c0}-1)^2}}
\left({ {\beta(\eta_{c0}-1)} \over {2 (2\nu^2+\alpha)}}\right)^{1/\gamma}
\left({{t}\over{t_c}} - {{(\eta_{c0}-1)}\over{\eta_{c0} \nu}}\right)
\right)^{{\gamma }\over{2\gamma-1}}.\label{aq}
\end{equation}
A photoevaporation timescale $t_{PE}$ for the clump can be defined
as the time for the radius of the clump to
shrink to zero, and from equation~(\ref{aq})
\begin{equation}
t_{PE}=t_c \left(\left({{r_s}\over{r_c(0)}}\right)^{2-1/\gamma}
\left({{3\gamma-1 }\over{2\gamma-1}}
\right){{(\eta_{c0}-1)^2}\over{6\nu \eta_{c0} }}\left({{2 (2\nu^2+\alpha)}
\over{\beta(\eta_{c0}-1)}}
\right)^{1/\gamma} + {{\eta_{c0}-1}\over{\eta_{c0}\nu}}\right).
\label{tpe2}
\end{equation}
For $\alpha=1, \beta=1$, and $\gamma=4/3$, the above equations determining 
the
evolution of the clump can be simplified as
\begin{equation}
r_c(t) = \left(\left({{r_s}\over{r_c(0)}}\right)^{5/4} - {{10 \eta_{c0}  \nu}\over{3
(\eta_{c0}-1)^2}}
q^{9/4} \left({{t}\over{t_c}} - {{(\eta_{c0}-1)}\over{\eta_{c0} \nu}}\right)
\right)^{4/5} r_c(0),
\end{equation}
\begin{equation}
m_c(t) = m_c(0) \left({{r_c(t)}\over{q r_c(0)}}\right)^{9/4},
\end{equation}
and
\begin{equation}
t_{PE}=t_c\left({{3(\eta_{c0}-1)^2}\over{10q\eta_{c0} \nu}}\left(1 -
{{3(1+q)}\over{\eta_{c0}-1}}
\right)^{5/9} + {{(\eta_{c0}-1)}\over{\eta_{c0} \nu}}\right).
\label{tpe2sim}
\end{equation}

\section{Evolution of clumps that undergo an initial expansion}
Clumps with large initial column densities and in low FUV fields quickly 
evolve
to the point where their internal pressures are greater than that of the
(expanded) thin heated surface layer. They expand out into vacuum (as they 
would
even in the absence of an external FUV field) at their sound speed $c_c$ for 
a
time $t_e$, until the pressure drops to that in the heated outer layer. At
$t=t_e$, there is pressure equilibrium and equation~(\ref{aj})  again holds. 
The clump expands to a new radius $r_c(t_e) = r_c(0) + c_c t_e$ and during 
expansion loses mass at a rate given by equation~(\ref{af}). 
The mass of the clump at $t=t_e$ is obtained as earlier,
\begin{equation}
m_c(t_e) = m_c(0) - 8  \pi m_H N_0 c_{PDR} (r_c(0) t_e + c_c
t_e^2/2).\label{ar}
\end{equation}
Also,
\begin{equation}
m_c(t_e) = {{4  \pi}\over {3}} m_H n_c(t_e) {r_c}^3 (t_e)^3. \label{as}
\end{equation}
Because there is significant expansion, $n_c(t_e)\ll n_{c0}$, we assume that
thermal pressure dominates in both the clump and the PDR flow. From
equations~(\ref{aj}) and~(\ref{as})
\begin{equation}
m_c(t_e) = {{8 \pi}\over {3}} m_H N_0  {{2\nu^2+ \alpha}\over{1+\alpha}}
r_c(t_e)^2 \label{at}
\end{equation}
Equations~(\ref{ar}) and~(\ref{at})   can be solved for $t_e$, and
\begin{equation}
t_e = t_c \left(1-{{1}\over{\eta_{c0}}}\right) \left(
\left( {{3\nu + \eta_{c0} -1}\over{3\nu +
2(2\nu^2+\alpha)/(1+\alpha)}}\right)^{1/2}
-1\right) \label{te}
\end{equation}

At times $t>t_e$, clump expansion is halted, and the clump is now confined
by the pressure at the base of the PDR flow. The clump slowly loses
mass, and shrinks to eventually become completely photoevaporated. The time
evolution of clump mass and size can be obtained by differentiating
equation~(\ref{at}) which also holds for $t>t_e$, with respect to time and
equating this with the mass loss rate (equation~\ref{ao} ). Thus,
\begin{equation}
r_c(t>t_e) = r_c(t_e) - \frac{3}{2} c_{PDR} {{1+\alpha}\over{2\nu^2+ \alpha}}(t-t_e),
\label{au}
\end{equation}
\begin{equation}
m_c(t>t_e) = m_{c0} {{2(2\nu^2+ \alpha)}\over{\eta_{c0}(1+\alpha)}}
  \left({{r_c(t)}\over {r_{c0}}}\right)^2.\label{av}
\end{equation}
The photoevaporation timescale is obtained by setting the clump radius to
zero at $t_{PE}$ to give
\begin{equation}
t_{PE} = \left({{2\nu^2+ \alpha}\over{1+\alpha}}\right)
{{r_c(0)+c_c t_e}\over{3 c_{PDR}}}
+ t_e, \label{tpeexp}
\end{equation}
and $t_e$ is given by equation~(\ref{te}).

\section{Numerical hydrodynamics code}
The fluid equations of motion for the system are solved using a 1-D 
spherical
Lagrangian hydrodynamics code. The equations are solved using a finite
difference method, with a numerical viscosity term, added as a 
pseudopressure,
for accurate handling of shocks and discontinuities (Richtmyer \& Morton
1967, Bowers  \& Wilson 1991). An isothermal equation of state is used
throughout.
The accuracy and stability of the scheme was checked with standard test
problems with known solutions. The computational grid has 1000 equally 
spaced
radial
zones, except for the central $20\%$, which was spaced logarithmically in
radius. This was done to increase spatial resolution at the centre, and it
provides
a more accurate calculation of any shock-induced collapse.

The FUV field is gradually turned on over a timescale, $t_{FUV}$, which is
varied
to accommodate both the impulsive and slow heating cases described in \S 4.
The temperature of an outer shell of material is thus raised continuously 
from
$T_c$ to $T_{PDR}$, reaching its maximum value of the PDR temperature at a 
time
$t_{FUV}$. The temperature $T$, exponentially drops off with the square of
the column, $N$, into the cloud, $T(N)=T_c+(T_{PDR}-T_c)e^{-(N/N_0)^2}$.
This profile  was chosen to closely mimic the $T$ dependence with column
predicted
by PDR models for the values of $G_0$ and $n$ under consideration (Kaufman 
et
al. 1999).

\end{document}